# Quantum Simulation of an Extended Fermi-Hubbard Model Using a 2D Lattice of Dopant-based Quantum Dots

Xiqiao Wang,[1,2] Ehsan Khatami,[3] Fan Fei,[1,4] Jonathan Wyrick,[1] Pradeep Namboodiri,[1] Ranjit Kashid,[1] Albert F. Rigosi,[1] Garnett Bryant,[1,2] Richard Silver[1*]

[1] Atom Based Device Group, National Institute of Standards and Technology, Gaithersburg, MD 20899

[2] Joint Quantum Institute, University of Maryland, College Park, 20740

[3] Department of Physics and Astronomy, San José State University, San José, CA 95192

[4] Department of Physics, University of Maryland, College Park, 20740

[*] Corresponding author

## Abstract

The Hubbard model is one of the primary models for understanding the essential many-body physics in condensed matter systems such as Mott insulators and cuprate high-$T_c$ superconductors. Due to the long-range Coulomb interactions, accessible low-temperatures, and atomic-scale nature, an artificial lattice of dopant-based quantum dots in silicon, consisting of one or a few dopant atoms per site, makes possible the analog quantum simulation of many-body problems that can be modeled by an extended Fermi-Hubbard Hamiltonian, particularly in the strong interaction regime. Effective control of tunable parameters in a dopant-based Hubbard simulator relies on the controlled placement of dopant atoms with atomic-scale precision. Recent advances in atomically precise fabrication in silicon using scanning tunneling microscopy (STM) have made possible atom-by-atom fabrication of single and few-dopant quantum dots and atomic-scale control of tunneling in dopant-based devices. However, the complex fabrication requirements of multi-component devices have meant that emulating two-dimensional (2D) Fermi-Hubbard physics using these systems has not been demonstrated. Here, we overcome these challenges by integrating the latest developments in atomic fabrication and demonstrate the analog quantum simulation of a 2D extended Fermi-Hubbard Hamiltonian using STM-fabricated 3×3 arrays of single/few-dopant quantum dots. We demonstrate low-temperature quantum transport and tuning of the electron ensemble using in-plane gates as efficient probes to characterize the many-body properties, such as charge addition, tunnel coupling, and the impact of disorder within the array. By controlling the array lattice constants with sub-nm precision, we demonstrate tuning of the hopping amplitude and long-range interactions and observe the finite-size analogue of a transition from Mott insulating to metallic behavior in the array. By increasing the measurement temperature, we simulate the effect of thermally activated hopping and Hubbard band formation in transport spectroscopy. We compare the analog quantum simulations with numerically simulated results to help understand the energy spectrum and resonant tunneling within the array. As atomically precise control of dopant-based quantum dots continues to improve, the results demonstrated in this study serve as a launching point for a new class of engineered artificial lattices to simulate the extended Fermi-Hubbard model of strongly correlated materials.

**Introduction**

Analog quantum simulators are designed quantum systems with a tunable Hamiltonian to emulate complex quantum systems intractable using classical computers due to the exponential growth of the Hilbert space with the system size.[1] Simulating strongly interacting fermions on a lattice lies at the heart of understanding quantum many-body phenomena, such as high-Tc superconductivity[2] and spin liquidity[3] that emerge in solid-state systems at low temperatures and are not describable through mean-field or density functional theory.

Various experimental platforms that form artificial lattices have been explored for realizing Fermi Hubbard analog quantum simulators, including optical lattices,[4,5] moiré superlattices,[6] and semiconductor quantum dot systems.[7,8] Quantum dots, often referred to as artificial atoms, can be arranged into artificial molecules and lattices with tunable hopping amplitude, interaction strength, and custom-designed point symmetry. For probing Fermionic many-body physics, the unique advantages of quantum-dot systems relative to other platforms, such as cold atoms and optical lattice systems, include readily achievable zero-temperature limit, easy access to transport measurements, and dynamic control of the chemical potential landscape and filling factors using gates.[9] Amongst the various semiconductor quantum dot systems, lattices of dopant-based quantum dots have unique advantages in simulating strongly correlated Fermionic systems of real atomic lattice sites because the atomic nature of the quantum dots means they have naturally occurring ion-cores, nuclear spins, hyperfine interactions, and inherently strong long-range interactions. Additionally, patterning the device geometry using the scanning tunneling microscope (STM)-based hydrogen lithography technique[10] adds the versatility of tailoring complex gate designs.

The Anderson-Mott transition has been previously demonstrated in few-atom systems using ion-implanted single dopant impurity atoms in silicon. However, the dopant positioning accuracy using the ion-implantation technique has been limited to ~30 - 50 nm.[11] Effective control of tunable Hamiltonian parameters and precision-engineering of electron and spin correlations in a dopant-based artificial lattice relies on the controlled placement of dopant atoms in the host lattice with near-atomic precision. Atomic-scale precision dopant placement can be achieved using the STM-based hydrogen lithography technique in silicon, initially pioneered by Joseph Lyding's group[12] and extensively developed by Michelle Simmons' group at UNSW,[10,13] and more recently by groups at NIST,[14,15] UCL,[16,17] and Sandia.[18,19] These advances have demonstrated success in atom-by-atom construction of single and few-dopant quantum dot devices in silicon and atomic-scale control of tunneling in dopant-based devices, enabling high-fidelity dopant-based multi-qubits.[20] (In this study we define "atomic-scale precision" as achieving sub-nm positioning accuracy of the articial lattice sites using single/few-dopant quantum dots.) Here we develop a path towards precise fabrication of dopant-based Fermi-Hubbard simulators where the on-site electron-electron interactions can be controlled by engineering the number and configuration of dopant atoms at each site, while the site-to-site hopping amplitude and long-range interactions can be controlled by altering the spatial separations between sites and electron occupation at each site. Limited theoretical studies have been carried out to predict the many-body properties in dopant-based quantum dot arrays in silicon, [21–23] such as strong correlation, excitation spectrum, and their robustness against disorder for analog quantum simulation. Transport through many-body states in dopant-based arrays has also been proposed as a sensitive probe to topological phase transitions[23] and coherent manipulation of electronics states.[24] Until now, however, atomic-scale fabrication and quantum transport characterization of 2D artificial lattices of single/few-dopants have not been realized in the laboratory.

Here, we demonstrate the quantum simulation of an extended 2D Fermi Hubbard Hamiltonian using atomic-scale fabricated 3x3 lattices of single/few-dopant quantum dots. We define the electron ensemble in the array by tuning the chemical potential landscape using in-plane gates and measure the low-temperature quantum transport through the array in a collective Coulomb blockade regime to probe many-body properties such as charge addition spectrum, resonant tunneling, and the impact of inhomogeneity within the array. By reducing the average lattice constant from ~10.7 nm to ~4.1 nm, we tune the site-to-site hopping amplitude and long-range interactions within the simulated Hamiltonian and observe the breakdown of the collective Coulomb blockade regime and a transition to metallic behavior. Finally, at elevated temperatures, we observe the formation of Hubbard bands in transport spectroscopy that can be attributed to additional hopping introduced by thermally activated occupation of many-body states. To augment the interpretation of the analog quantum simulations, we numerically solve the Hubbard Hamiltonian using exact diagonalization and parameters estimated based on the experimental characterization of the array. Our results establish a new solid-state platform for the quantum simulation of extended Fermi-Hubbard models of strongly correlated 2D systems.

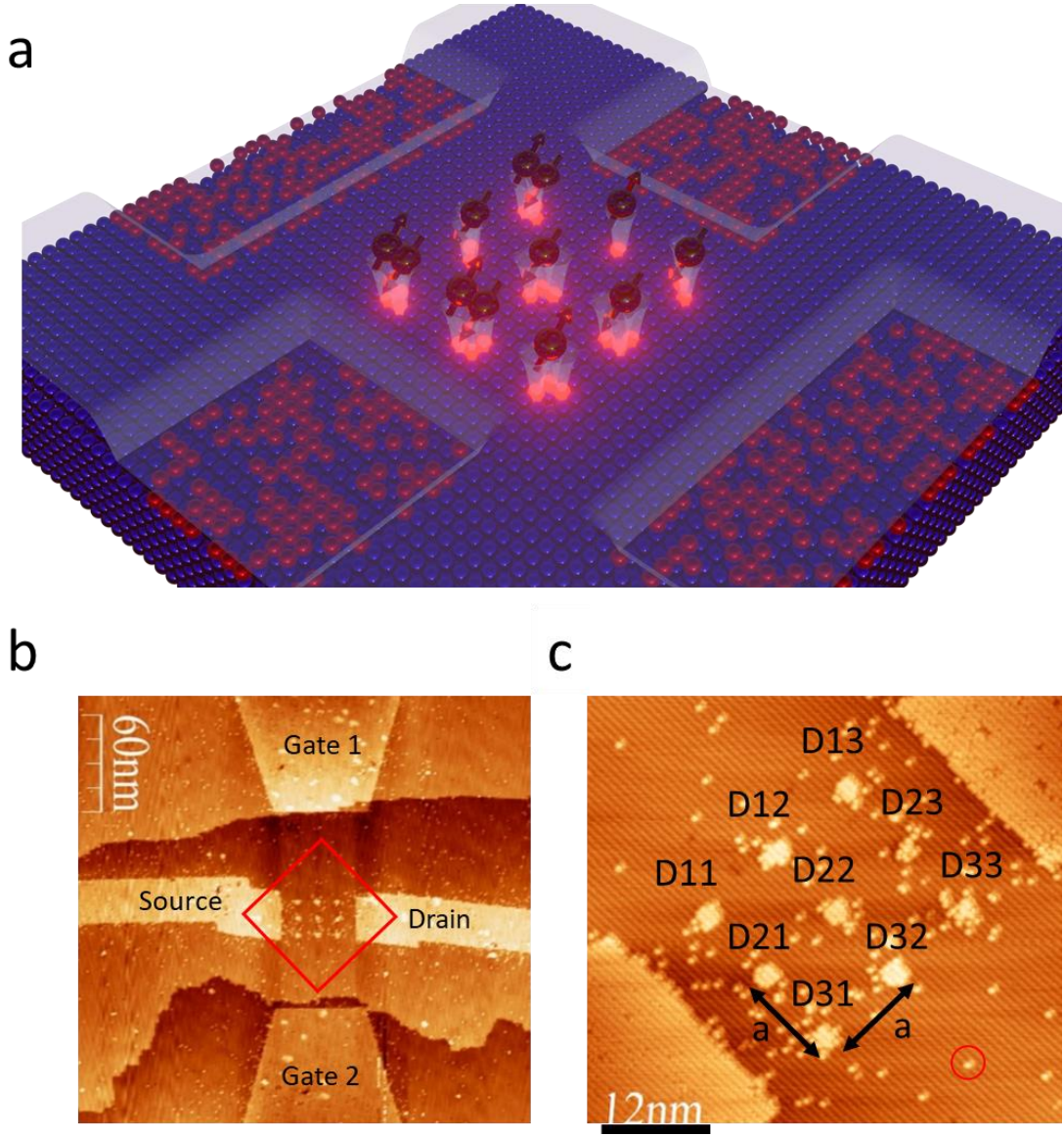

Figure 1. 2D dopant-based quantum dot arrays as a platform for simulating the extended Fermi Hubbard model. (a) Schematic of the experimental Fermi-Hubbard system composed of a 3x3 array of single/few-dopant quantum dots coupled to in-plane gates and source-drain leads, allowing transport measurements through the array. (b) STM image of the central device region of the 3x3 array acquired immediately following hydrogen lithography. The lithography patterns appear as bright regions on the hydrogen-terminated Si(100) surface due to hydrogen-depassivation and the exposure of chemically reactive Si dangling bonds. In this image, the central array and the source/drain leads reside on the same Si(100) surface terrace; Gate 1 and Gate 2 reside on surface terraces one mono-atomic layer (~0.14nm) above and below the middle terrace. (c) Atomic resolution STM image of the 3x3 array pattern (zoom in of marked square region in (b)). Each dot is numbered to facilitate the discussion in the main text. In this array, $a = 10.7 \pm 0.3\ nm$ is the square lattice constant that equals the average center-to-center distance between nearest neighboring sites. 2×1 surface reconstruction dimer rows on the Si(100) surface run from the upper left to lower right direction in the image. We define the 2D

square lattice constant, a, as the averaged center-to-center distance between nearest neighboring dots within the array. The red circle marks an example of an isolated single dangling bond that does not incorporate dopant atoms. The STM image is taken at −2 V sample bias and 0.1 nA setpoint current.

## Results

### 3x3 Arrays of Few-dopant Quantum Dots and the Extended Fermi-Hubbard Hamiltonian

We fabricated[14] a series of 3x3 square lattice arrays of few-dopant quantum dots that are weakly tunnel coupled to a source and drain and capacitively coupled to two in-plane gates. Figure 1 shows STM images of the hydrogen lithography patterns from one of the arrays (average lattice constant $a_1 = 10.7 \pm 0.3\ nm$) on a hydrogen-terminated Si(100) 2×1 reconstruction surface, where the locations of the artificial lattice sites and the lattice constants can be determined by using the surface reconstruction dimer unit cells as an atomically precise ruler and counting the number of dimer rows (dimer-row pitch = 0.77nm) between neighboring sites. Each lattice site is defined by using an STM tip to remove a small patch of (~10 to ~20) adjacent hydrogen atoms, allowing individual phosphorus atoms to incorporate only into the exposed surface Si lattice sites in a subsequent phosphorus dosing and incorporation process. We have recently demonstrated that a dangling bond patch of similar size typically forms a few-dopant cluster quantum dot incorporating 1 to 3 phosphorus atoms.[14] At the same time, the STM-patterned in-plane source/drain leads and two symmetric in-plane gates are saturation-doped to a dopant density of ~$2 \times 10^{14}/cm^2$ that corresponds to a bulk doping density of $2 \times 10^{21}/cm^3$, approximately three orders of magnitude above the bulk metal-insulator transition, allowing quasi-metallic conduction in all electrodes. We use a room-temperature grown locking layer technique to suppress atomic-scale movement of the precisely defined dopant atom positions [25] before a subsequent low-temperature (~250 ℃) epitaxial Si overgrowth, that embeds the dopant atoms in a 3-dimensional crystalline Si environment. Finally, ohmic contacts to the buried electrodes are formed using a low thermal budget silicide contacting technique.[26]

The Hubbard model has long provided a theoretical playground for understanding different phases of matter, especially in the presence of strong electronic correlations, where conventional density functional theory (e.g. employing the local density approximation functional) fails. The 3x3 array's physical attributes map to an extended Fermi-Hubbard Hamiltonian, which, in its simplest form, includes one spinful orbital at each site. Experimentally, the absolute number of excess electrons at each single/few-dopant quantum dot may vary according to the inhomogeneity in the number of dopant atoms at each lattice site. For quantum transport through dopant-based quantum dots within the small bias ranges used in this study, charge fluctuations at quantum dots occur via quantum dot energy levels that are near or in between the Fermi levels in the source and drain leads. Therefore, we limit our analysis to on-site binding energy levels that are nearest to the Fermi level, i.e., charge number fluctuations of up to two electrons at each site which corresponds to the three charge states of a few-dopant quantum dot: the ionized D⁺ state, the charge-neutral D⁰ state, and the negatively charged D⁻ state. The absolute number of excess electrons on a few-dopant quantum dot does not affect the underlying physical phenomena in this work.

$$H = H_\mu + H_t + H_U$$

$$H_\mu = \sum_{i\sigma} \mu_i n_{i\sigma} = \sum_{i\sigma}\left(p_i + E_{bi} + \sum_{j, i \neq j} V_{i,j}\right) n_{i\sigma}$$

$$H_t = \sum_{\langle i,j \rangle, \sigma} \left( -t c_{i\sigma}^\dagger c_{j\sigma} + H.c. \right)$$

$$H_U = \sum_i U_i n_{i\uparrow} n_{i\downarrow} + \sum_{i,j, i \neq j} U_{i,j} n_i n_j$$

Equations 1.

The total Hamiltonian consists of the onsite energy terms in $H_\mu$, the kinetic energy (hopping) terms in $H_t$, and the electron-electron interaction energy terms in $H_U$. Here, $n_{i\sigma} = c_{i\sigma}^\dagger c_{j\sigma}$ is the number operator where $c_{i\sigma}^\dagger$ ($c_{i\sigma}$) is the creation (annihilation) operator of a fermion with spin $\sigma$ at lattice site $i$. $\mu_i$ is the chemical potential at site $i$ comprised of fixed contributions from local and long-range electron-ion core Coulomb interactions, i.e., $E_{bi}$ (electron binding energy at site $i$) and $V_{i,j}$ (Coulomb attraction between an electron at site $i$ and an ion-core at site $j$); $E_{bi}$ is determined by the number of dopants and detailed dopant-cluster configurations at each site. $V_{i,j}$ is determined by the separation between two lattice sites and can be approximated using a point-charge approximation $V_{i,j} \approx -\frac{V_0}{|R_i - R_j|}$ where $V_0 = \frac{e^2}{4\pi \varepsilon_r \varepsilon_0} \approx 123 meV \cdot nm$ in silicon (see Supplementary Materials Section S1). $\mu_i$ also includes a tunable contribution $p_i$ that is determined by the classical capacitance couplings of the device and the applied voltages on the gates and source/drain leads. $t$ represents the hopping amplitude between adjacent sites ($H.c.$ indicates Hermitian conjugate). $U_i$ is the local electron-electron Coulomb repulsion at site $i$, and $U_{ij}$ is the long-range Coulomb repulsion between electrons at sites $i$ and $j$. The filling factor (electron number) in the array is determined by the position of array's chemical potential with respect to the Fermi levels in the source/drain leads, which we set as $E_F = 80 meV$ below the Si conduction band edge. [10] We ignored Coulomb exchange and higher-order hopping terms in our numerical model.

The hopping $t$, interactions $U_i$, $U_{i,j}$, chemical potential terms $\mu_i$, which determines the electron ensemble (electron distribution and doping level of the array), and the temperature $T$ constitute the Fermi Hubbard model's parameter space, covering a large variety of condensed matter systems, some not yet fully understood, and others that continue to be discovered. Physical control of the Hubbard model parameters in our 2D arrays is achieved by varying device fabrication and measurement conditions. The number of electrons in the array ($\sum_i n_i$), which determines the filling factor, can be altered by applying a common voltage on both in-plane gates to sweep the array's global chemical potential with respect to the Fermi level. A voltage difference between the two gates will introduce a potential gradient within the 2D array along the gate-gate direction. In contrast to gate-defined quantum dot systems, in-situ gate tuning of tunnel coupling within a single device is less efficient in donor-defined quantum dots systems. Effects of tuning the hopping amplitude $t$ and long-range interaction terms ($U_{i,j}$) can be achieved, however, by fabricating a series of dopant-based lattices with different lattice constants that are determined at the fabrication stage. The local interaction term $U_i$ reflects the physical size of each quantum dot. For the arrays in this study, we design the in-plane gates and average lithographic dot size the same for all arrays and only alter the lattice constant from $a_1 = 10.7 \pm 0.3 nm$ in the first array to $a_2 = 6.6 \pm 0.3 nm$ in the second array and to $a_3 = 4.1 \pm 0.3 nm$ in the third array. Based on previous theoretical studies,[27,28] these lattice constants correspond to hopping amplitudes in the hundreds of $\mu eV$ to the few-$meV$ range and long-range interactions in the range of few-$meV$ to ~ 20meV (See Supplementary Materials Section S1). We estimate the number of dopant atoms in the quantum dots to be within the range of $2 \pm 1$ dopants by characterizing the binding energies of few-dopant quantum dots with a similar lithographic patch (See Supplementary Materials

Sections S2 and S3). A quantum dot size of $2 \pm 1$ dopants corresponds to a local electron-electron interaction energy $U_i$ of ~45meV (for the relevant $D^0$ to $D^-$ transitions). These energy scales position these Hubbard arrays in the strongly interacting regime with non-negligible long-range interactions. As also supported by our theory calculations, the latter are, however, not large enough to result in the charge density wave ground state.[29]

Atomic-scale defects have been a critical challenge for solid-state implementations of quantum devices that rely on atomic precision fabrication processes such as those used here. The primary source of defect/disorder is the site-by-site variation in on-site energies $E_b$ and $U_i$, which effectively introduces inhomogeneity in the chemical potential landscape, interactions, and hopping amplitudes. Due to the stochastic nature of the phosphorus dosing and incorporation process,[30] deterministic control of the exact number of dopant atoms and their specific cluster configuration within a lithographic patch remains an unsolved challenge in the community.[18] While the precise atomic configurations in an array may, in principle, be obtained by parametrically fitting the measurement results with numerical simulations, a detailed disorder configuration investigation is extremely computationally expensive and beyond the scope of this study. Instead, we account for the effects of disorder by estimating the number of dopant atoms-per-site, based on STM-lithography patterns and dopant incorporation conditions,[31] for use as input to the numerical simulations. While we have not pursued the exact match between the numerically simulated and experimentally simulated results in this study, the detailed atomic configuration at each array site does not alter the qualitative understanding of the array system (see Supplementary Materials Section S5), and the quantitative differences between theory and experiment are a measure of the accuracy of the disorder estimates. Table S3 in the Supplementary Materials lists the range of binding energy $E_b$ and the on-site addition-energy $U_i$ for 1P, 2P, and 3P dopant clusters.[28] Variations in the number of dopant atoms and detailed dopant cluster configurations at each lattice site introduce site disorder, which in larger arrays lends itself to the study of Anderson localization, and especially its fate in the presence of strong interactions.[27]

In the following sections, we demonstrate the tunability of dopant-based 2D arrays for analog quantum simulation by first demonstrating the ability to define the electron ensemble and characterize the charge addition boundaries and resonant tunneling within the 2D array using in-plane gates. Then we show tuning of the hopping amplitudes and long-range interactions by changing the lattice constants within the array. Finally, we measure the transport spectrum at elevated temperatures and reveal thermally activated Hubbard bands within the array.

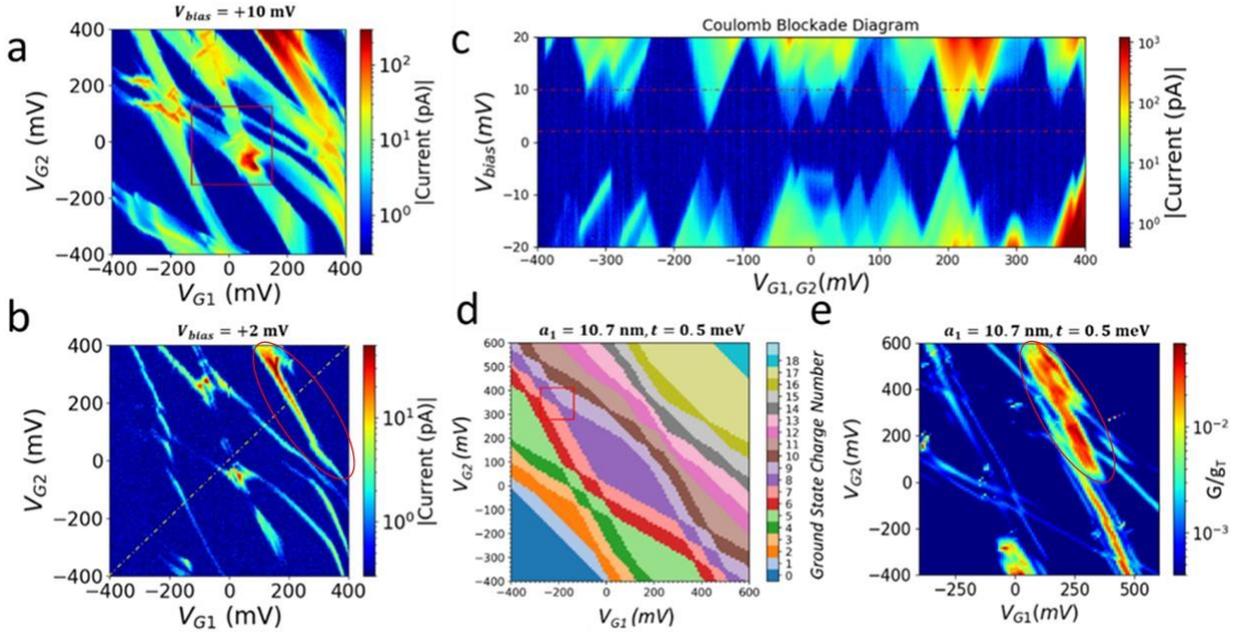

Figure 2. Conductance spectroscopy and charge stability diagram of the 3×3 array on a square lattice shown in Figure 1. (a, b) Experimental conductance gate-gate maps measured at base temperature (T~10mK) by applying +10mV bias (a) and +2mV bias (b) to the drain-lead while grounding the source lead and monitoring the drain-source direct current (DC) while independently sweeping the gate voltages at the upper and lower in-plane gates. (c) Finite bias conductance spectroscopy taken from a diagonal cut along the dashed line in (b). Horizontal dashed lines in (c) correspond to the bias voltage levels in (a) and (b). (d) Numerically simulated charge stability diagram of the 3×3 array's ground state (see Methods). The hopping amplitude is set as t = 0.5meV. (e) Numerically simulated resonant conductance gate-gate map at $T = 1K$ (see Methods).

### Gate-tuning the Electron Ensemble and Charge Distributions

At a base temperature of T~10mK, we measure the transport spectrum through the array by monitoring the source-drain direct current (DC) while sweeping gate-gate voltages adiabatically at finite bias; see examples of conductance gate-gate maps (charge stability diagrams) in Figs. 2a and 2b, or sweeping source-drain bias adiabatically at a given set of gate voltages; see the bias conductance spectrum (Coulomb diamond plots) in Fig. 2c taken from a diagonal cut along the dashed line in (b). Due to the weak tunnel coupling between the array and source/drain leads and finite local and long-range electron-electron interactions, the source-drain conductance through the array is in the collective Coulomb blockade transport regime [33,34] as evidenced by the zero-conductance regions at finite biases in the measured conductance maps. The collective Coulomb blockade can be lifted by applying plunger gate voltages that overcome the interaction-induced blockade barrier and align an addition energy level of the array with the Fermi level in the source and drain leads. At these gate conditions the electron number in the array can fluctuate by one (finite compressibility), allowing source-drain conductance through the array.

Figs. 2d and 2e plot numerically simulated charge stability diagrams of the ground states and conductance map over the gate-gate space, respectively. Within each color domain in the simulated charge stability diagram, the total number of excess charges in the array ($N$) is constant, corresponding to the Coulomb blockaded regions in transport. Sweeping the common-voltage applied to both gates along the 45-degree diagonal direction on the charge stability diagram controls the total number of electrons, therefore, the filling factor of the array. Sweeping the differential voltage between the two in-plane gates (along the 135-degree diagonal direction in Fig. 2d) effectively tilts the chemical potential landscape within the array, altering the charge distribution of the ground state without considerably affecting the filling factor within the array. Due to variations in the input binding energies (see Supplementary Materials Section S2), the largest Coulomb blockaded region in the theory diagram belong to N=8, corresponding to ~11% hole doping, as opposed to half filling (N=9), which is expected for the uniform Fermi-Hubbard model.

The charge stability domain boundaries correspond to conductance lines in transport maps. Comparing the simulated conductance map with the charge stability diagram, not all charge addition boundaries are equally visible in conductance. This is also evident in the non-closing diamonds in Fig. 2c.[35] Both the measured and simulated conductance maps agree on a relatively large conductance along a particular line of negative slope in the positive gate voltages region of the diagram (circled in Figs. 2b and 2e). Visibility of the charge addition boundaries in measured conductance maps can be enhanced by increasing the bias window (See Figs. 2a and 2b) to allow transport through excited states or other inelastic/incoherent transport processes in the array mediated by, for example, electron-phonon interactions. A 2D array can allow a combination of serial and parallel transport configurations through the dot array between the source-drain leads. We attribute the non-symmetric shape of the charge stability diagrams with respect to two gate voltages, or the fact that Coulomb diamonds do not always close (do not have conductance) at a small bias along the diagonal gate-gate direction, to the following: First, we expect the dominant effect to be the non-uniformity in binding energies due to variations in the number of dopants per site, which not only makes the system's energy landscape non-uniform for the electrons but also affects the transport through the array. Additional effects include small asymmetries in the way each gate affects individual site potentials and atomic-scale variation in the hopping amplitudes.

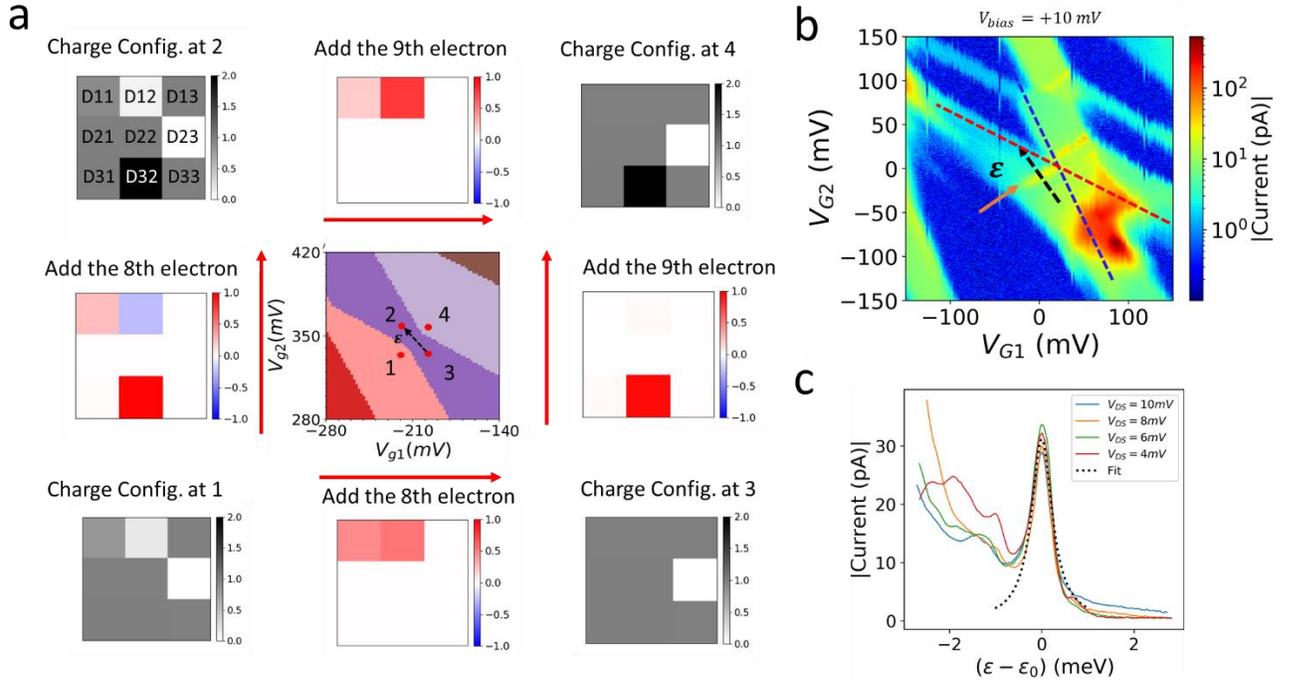

Figure 3. Charge addition and resonant tunneling at avoided crossings in the conductance gate-gate map. (a) Schematic illustration of eigenstate charge distributions and single charge addition at a charge stability diagram region as highlighted by the red box in Fig. 2d. The dot numbering scheme from Fig. 1c is overlaid on the first charge configuration panel. The black-white color maps represent the charge occupation of the ground states at select locations. The red-blue color maps represent the changes in ground state occupation when crossing a charge stability boundary and adding a single electron onto the array. The detuning axis $\varepsilon$ between eigenstates at points 2 and 3 is marked by the dashed arrow. (b) Close-up of a region containing a few avoided crossings as highlighted by the red box in Fig. 2a. The solid arrow indicates a resonant conductance line. The dashed lines mark the slopes of the charge addition conductance lines that come across near the resonant transitions. The dashed arrow marks the gate detuning axis perpendicular to the resonant conductance line. (c) Overlayed resonant conductance profiles along the gate detuning axis that are measured at different biases. Each curve is fitted using Equation 2 in the main text; and the best-fit parameters are averaged to reconstruct the dotted curves.

With our symmetric, in-plane two-gate design, the ratio of the upper gate's (Gate1) and lower gate's (Gate2) capacitive coupling to electrons at a quantum dot is determined by the ratio of the quantum dot's distances to the upper gate and to the lower gate. Charge additions to quantum dots in the upper, middle, and lower rows in the array correspond to three distinct lever arm ratios, which manifest in the charge stability diagram as three different slopes in charge addition boundaries. To illustrate this, we zoom in at an avoided crossing at $N = 8$ in the simulated charge stability diagram in Fig. 3a and plot the ground state charge distributions and charge addition configurations at select gate-gate positions. When gating the array from positions 1 to 2 and from 3 to 4, we cross charge addition boundaries of the same slope, corresponding to the addition of an electron onto sites in the lower row. Similarly, when gating the array from positions 1 to 3 and from 2 to 4, the charge addition boundary slope corresponds to adding an electron onto sites in the upper row. Charge addition to the middle row can be found elsewhere in Fig. 2d when crossing addition boundaries with an intermediate magnitude in slope.

Additionally, we find small resonant tunneling lines with positive slopes in the conductance map, which we attribute to hybridization between many-body states with the same filling factor but different charge distributions on the array. Because of the 'multiply connected' topology[36] in the 2-D array, the resonance conditions within the array enhance conductance through the array by opening additional conduction paths. Examples are ground states at positions 2 and 3 in Fig. 3a with an avoided crossing of charge addition lines between them. The slopes of the two crossing charge addition lines (dashed lines in Fig. 3b) correspond to charge addition to sites in the upper row and to sites in the lower row, respectively. Approximating the two ground states as classical states of a double-dot system in which the electron is localized at either the left or the right dot, we can find an estimate for the effective tunneling between the states, t', by fitting the conductance peak at the avoided crossing to the following Lorentzian form[37,38]

$$\frac{I_{ds}}{e} = \frac{t'^2 \Gamma_d}{t'^2 \left(2 + \frac{\Gamma_d}{\Gamma_s}\right) + \frac{\Gamma_d^2}{4} + \frac{1}{\hbar^2}(\varepsilon - \varepsilon_0)^2}$$

Equation 2

where $I_{ds}$ is the drain-source current; $t'$ is the tunnel coupling rate ($1 GHz \approx 4.14 \mu eV$) between two charge occupations of discrete energy levels. $\Gamma_d$ and $\Gamma_s$ are tunneling rates to or from the drain and source leads, which are assumed to be approximately equal in this study; $(\varepsilon - \varepsilon_0)$ is the detuning energy between the two discrete energy levels (see black dashed arrow in Fig. 3b, based on $V_{G1}$ and $V_{G2}$ using lever arms, see Supplementary Materials Section S1); and $h$ and $e$ are Plank's constant and the charge of a single electron. The amplitude of the current peak is primarily determined by $\Gamma_d$ and $\Gamma_s$ ; and the width of the peak depends mostly on $t'$. We treat $t'$, $\Gamma_d$, and $\Gamma_s$ as fitting parameters.

As shown in Figs. 3b and 3c, we perform the fit using the measured conductance near a crossing along the black dashed line in Fig. 3b. We find $t' \approx 4.4 \pm 0.3 GHz (14.1 \pm 1.7 \mu eV)$ and $\Gamma_d \approx \Gamma_s \approx 0.6 \pm 0.1 GHz$. The independence of the resonant current peak height and peak shape on the applied bias confirms that the observed peaks are dominated by resonant transport through two discrete energy levels. Away from zero-detuning, the two discrete energy levels are isolated, while at zero-detuning, the two discrete energy levels maximally hybridize. For a double-dot system, the separation at an avoided crossing is determined by $(U_m + 2t')$, where $U_m$ is the long-range e-e repulsion between the two sites. We computationally verify that this relationship also applies to our 3x3 array systems where two localized charge distribution configurations that are on resonance at an avoided crossing can be mapped to the two sites of a double-dot system. (See Supplementary Materials Section S6) We find that the magnitude of $t'$ from the above fit corresponds to a simulated tunnel coupling that is between non-neighboring arrays sites (See Supplementary Materials Table S4), which is consistent with the slopes of the crossing charge addition lines that correspond to sites in non-adjacent (upper and lower) rows. Here each $t'$ should be interpreted as an off-diagonal matrix element for an effective two-level Hamiltonian near the avoided crossing, which should be distinguished from the nearest neighbor hopping amplitude $t$ in the Fermi-Hubbard Hamiltonian.

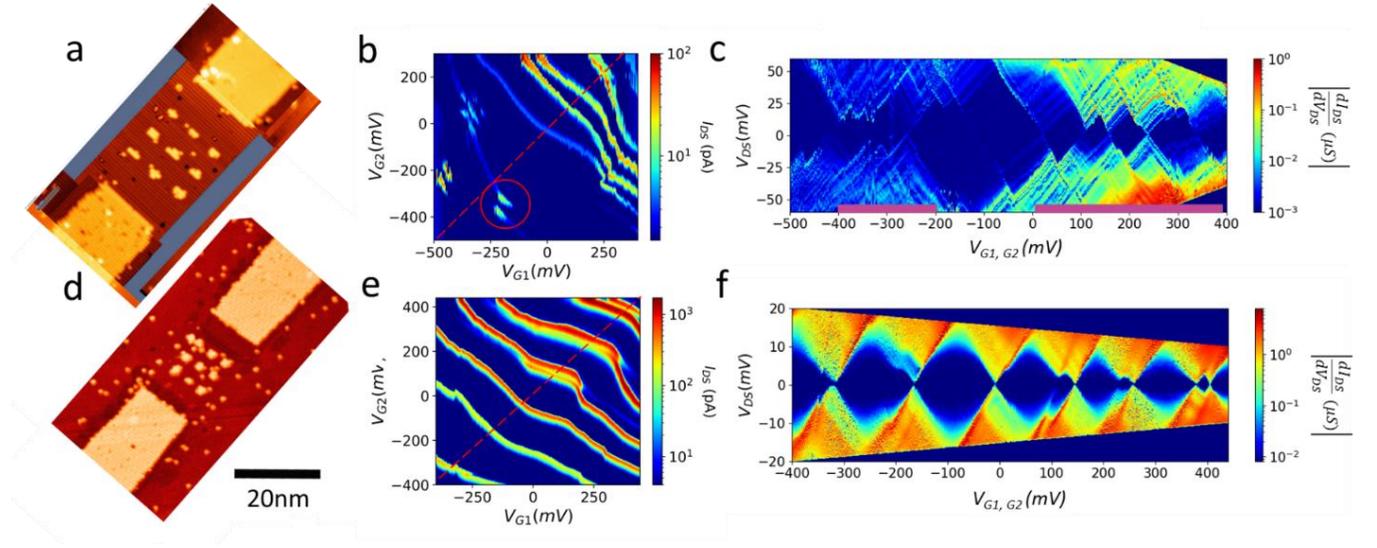

Figure 4. Transition to more metallic behavior at reduced lattice constant. (a) (d) STM images of the second and the third array where the lattice constants are $a_2 = 6.6 \pm 0.3 nm$ and $a_3 = 4.1 \pm 0.3 nm$, respectively. (b) (e) Experimentally measured DC conductance charge stability diagrams of the second array (b) at $V_{bias} = 6mV$ and the third array (e) at $V_{bias} = 2mV$ at the base temperature of the dilution refrigerator ($T = 10mK$). (c) (f) Differential conductance Coulomb blockade diagrams of the second and the third arrays that are measured at $T = 10mK$ along the dashed lines in (b) and (e), respectively.

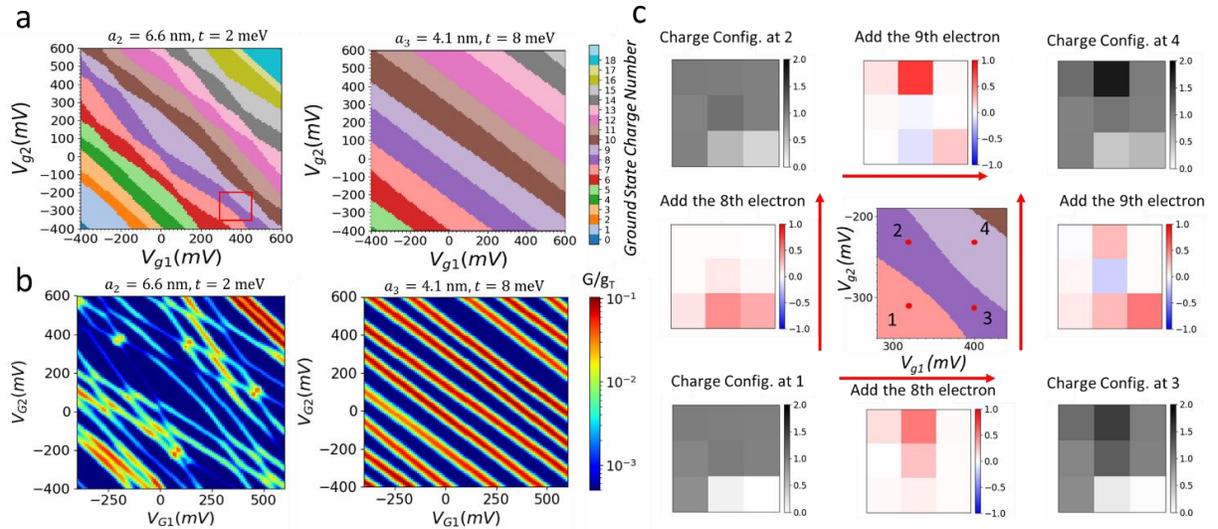

Figure 5. Effects from decreasing the lattice constants within the array. (a) Numerically simulated charge stability diagrams with decreased lattice constants and increased hopping amplitudes. (b) Numerically simulated conductance maps of the corresponding charge stability diagram in (a) at $T = 1K$. (See Methods) Both sets of diagrams in (a) and (b) share the same color bars on the right. (c) Schematic illustration of eigenstate charge distributions and single charge addition at a charge stability diagram region as highlighted by the red box in the left diagram in (a). Note the significant increase in electron

delocalization as the lattice constants are reduced. The ground state charge distribution and charge addition plots following the same convention as described in Fig. 3a.

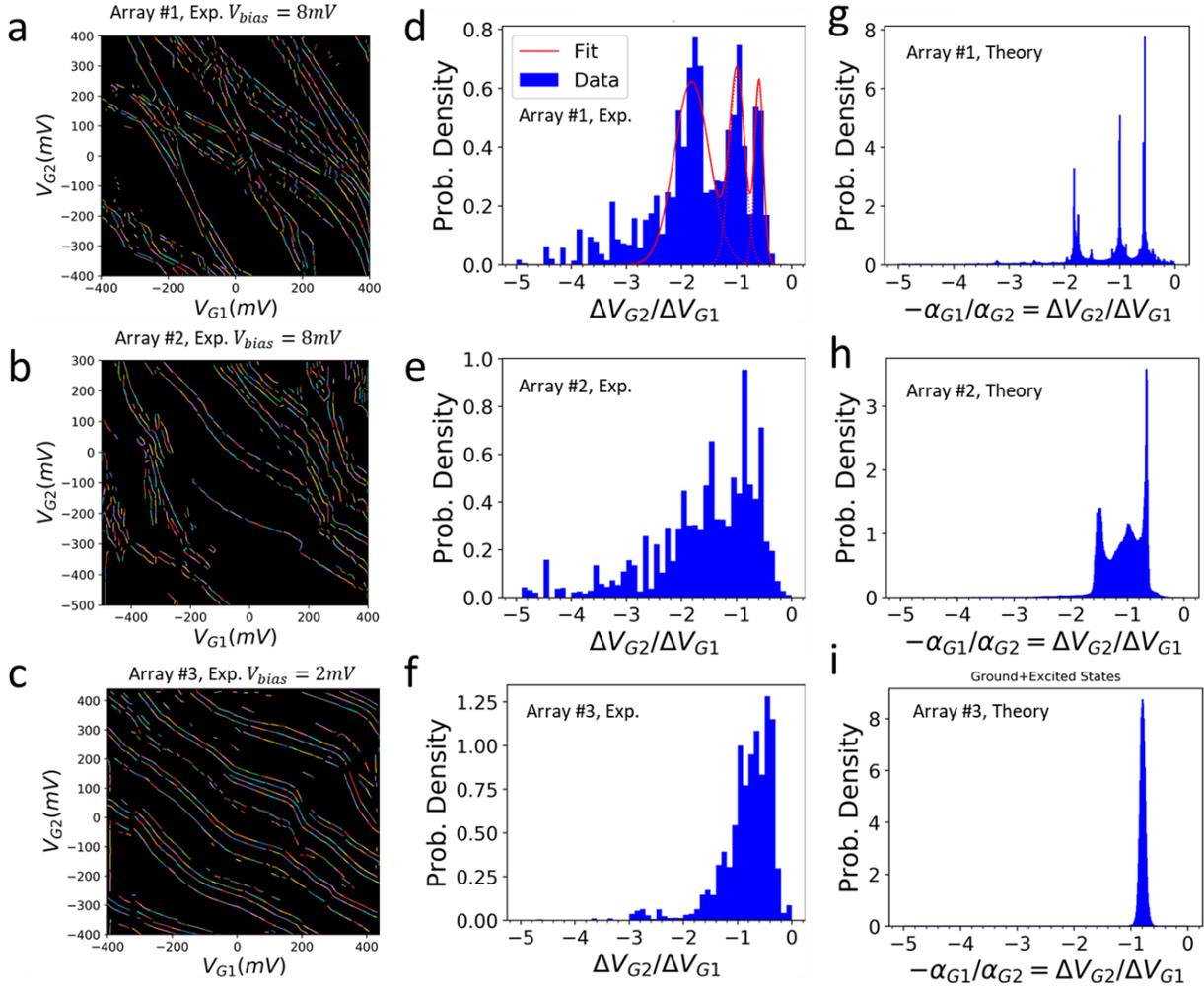

Figure 6. Histogram distributions of conductance features in gate-gate maps for all three arrays. The slopes of the conductance lines contain spatial information about the many-body states through which addition electron transport ocurs. (a) (b) (c) Edge detection of the conductance features in the gate-gate maps of the first array at $V_{bias} = 8mV$ (a), the second array at $V_{bias} = 8mV$ (b), and the third array at $V_{bias} = 2mV$ (c), respectively. The edge detection is performed using a Hough transform. The detected edges of constant slopes are represented by colored line segments. We quantify the span of each line segment by measuring the length of the line segment in units of pixel in the gate-gate map. (d) (e) (f) Histogram distributions of the detected conductance line slopes of the first (d), the second (e), and the third (f) arrays, respectively. The histograms are plotted as probability density distributions of the number of pixels of the detected conductance line segments as a function of their slopes from the gate-gate maps. The three histogram peaks in (d) are fitted using Gaussian functions. (g) (h) (i) The calculated histogram distributions of the gate lever arm ratio of electron addition occupancies over the simulated charge stability maps of the first array (g), the second (h), and the third (i) arrays, respectively. The calculated histogram distributions include states within the corresponding bias windows in (a), (b), and (c).

## Tuning the Hopping Amplitude and Long-range Interactions

The nearest neighbor hopping $t$ is exponentially dependent on the lattice-constant $a$ while the long-range e-e Coulomb interactions are inversely proportional to $a$. Experimentally, we tune the hopping amplitude and long-range interactions in the 3x3 arrays by reducing the lattice constant $a_1 = 10.7 \pm 0.3 nm$ in the first array (shown in Fig. 1c), to $a_2 = 6.6 \pm 0.3 nm$ in a second array, and to $a_3 = 4.1 \pm 0.3 nm$ in a third array, see Fig. 4. If we view each artificial lattice site as an 'impurity' in silicon, the lattice constant in the first, second, and the third arrays correspond to a bulk doping density of $\sim 8 \times 10^{17}/cm^3$, $\sim 3.5 \times 10^{18}/cm^3$, and $\sim 1.5 \times 10^{19}/cm^3$, spanning from below to above the critical density of a metal-insulator transition in phosphorus-doped bulk silicon.

Figs. 4b and 4e are the measured charge stability diagrams from the second and the third arrays, respectively. Compared with the charge stability diagrams of the first array in Fig. 2, a key observation is that the charge stability boundaries in the first array are dominated by straight line segments of distinct negative slopes with bias triangles at avoided crossings; whereas on the gate-gate plane in the second array, the charge addition boundaries appear crossed only at negative gate voltages and evolve into smooth curves at positive gate voltages. In the third array, charge addition boundaries run more or less parallel throughout the entire gate-gate plane, resembling the charge stability diagram of a single island in a single-electron transistor (SET). This is further substantiated by comparing the differential conductance bias spectra of the second and third arrays (Figs. 4c and 4f) measured at T~10mK base temperature along the dashed lines in Figs. 4b and 4e, respectively. Here the Coulomb diamonds in the third array (Fig.4f) are dominated by well-defined diamond shapes that are closing at small biases and of similar diamond heights (charging energy), resembling the conventional Coulomb blockade behavior of a metallic island SET. However, despite the dominant metallic behavior in the gate-gate map, we do observe small discontinuities in the transport lines in the higher voltage range, likely due nearby charge instabilities or residual disorder effects. In contrast, the Coulomb diamonds of the second array (Figs. 4c) are irregularly shaped, and, despite the expected presence of disorder within the arrays, the bias spectrum of the second array features two groups of small Coulomb diamonds (highlighted by magenta color bars), corresponding to the upper and lower Hubbard bands, separated by a set of large, irregularly shaped Coulomb diamonds, corresponding to the Mott gap. The separation of the two Hubbard bands is characterized by the heights (addition energy) of the Coulomb diamonds at the Mott gap, ~50meV, which is in quantitative agreement with the addition-energy of individual quantum dots of few (1 to 3) dopant atoms. [28] (See Supplementary Materials Section S3) Additionally, in Fig. 4c and 4f, we observe a qualitative difference in the bias conductance spectrum between the two arrays at base temperature. In the second array (Fig. 4c), resonance conductances are visible as lines of increased differential conductance running parallel to the edges of Coulomb diamonds, indicating a discrete eigen-enegy spectrum within the array; such a discrete conductance spectrum is not visible in the third array (Fig. 4f), indicating a quasi-continuous (metallic) density of state distribution in the third array.

As discussed above, avoided crossing separations in charge stability diagrams are determined by the effective mutual charging energies $U_m$ and tunnel coupling $t'$ between relevant many-body states. Figs. 5a and 5b plot the simulated charge stability and conductance diagrams to illustrate the effects of reduced lattice constants and increased hopping amplitudes in the second and the third arrays. For the left and the right diagrams in Figs. 5a and 5b, the input parameters of capacitance matrices and estimates of the numbers of dopants per site are based on the device geometries of the second and the third arrays, respectively (see Fig. 4a and 4d, and Supplementary Materials Sections S1 and S2); and the hopping amplitudes are set at t=2meV and t=8meV, respectively. Compared with the simulated diagrams for the first array in Figs. 2d and 2e, the simulated diagrams for the second array (left panels in Figs. 5a and 5b) feature the expected larger avoided crossing separations, which, in the measured charge stability diagram of the second array (Fig. 4b), correspond to the increased separations within bias-triangle pairs as compared to those in the first array (Figs. 2a and 2b). To illustrate the delocalization of many-body states in the second array, Fig. 5c plots ground state charge and charge addition distributions at select positions near an avoided crossing (highlighted by the red square in the left panel in Fig. 5a). Here, the addition electrons are clearly more delocalized than those shown in Fig. 3a for the first array. As $U_m$ and $t'$ are further increased for the simulations of the third array (right panels in Figs. 5a and 5b), the array behaves like a single island and appears less sensitive to detailed disorder configurations within the array, giving better agreement between the simulated conductance map and the measured conductance map (Fig. 4e) for the third array as compared with the results of the first and the second arrays.

We carry out a quantitative analysis by running edge detection algorithms on the measured charge stability maps of the three arrays (See Figs. 6a, 6b, and 6c) and plotting histograms of the detected edge length along their negative slopes $\frac{-\Delta V_{G2}}{\Delta V_{G1}}$. As can be seen in Figs. 6d, 6e, and 6f, the histogram of the first array shows three narrow distinct peaks, corresponding to charge addition sites in the upper, middle, and lower rows, and little mixing occurs in states between different rows for conductance via single electrons jumping on and off the array. In the second array (Fig. 6e), in contrast, the three distinct peaks in the distribution evolve into a single broad distribution, exhibiting conductance that occurs via single addition electrons that are more often shared simultaneously by sites across different rows. In the third array (Fig. 6f), the single peak in the distribution narrows down further, mimicking the behavior of conductance through a metallic SET island. Unstructured distributions at higher slopes ($\Delta V_{G2}/\Delta V_{G1}$<-3) are likely due to combined effects of defects and edge detection artifacts. Similar results are found in Figs. 6g, 6h, and 6i plotting the histogram distributions of the lever arm ratios of charge additions, which are calculated from the simulated charge stability diagrams using the identical bias windows as in the experimental plots. (see Supplementary Materials Section S1 for detailed capacitance and lever arm calculation methods). This is characteristic of the melting of the collective Coulomb blockade regime upon increased hopping amplitude relative to the interaction strength. Such a transition from the collective Coulomb blockade regime to more metallic behavior within the array is also observed in Figs. 4c and 4f as the many-body energy spectrum transitions from discrete to quasi-continuous distributions in the second and the third arrays.

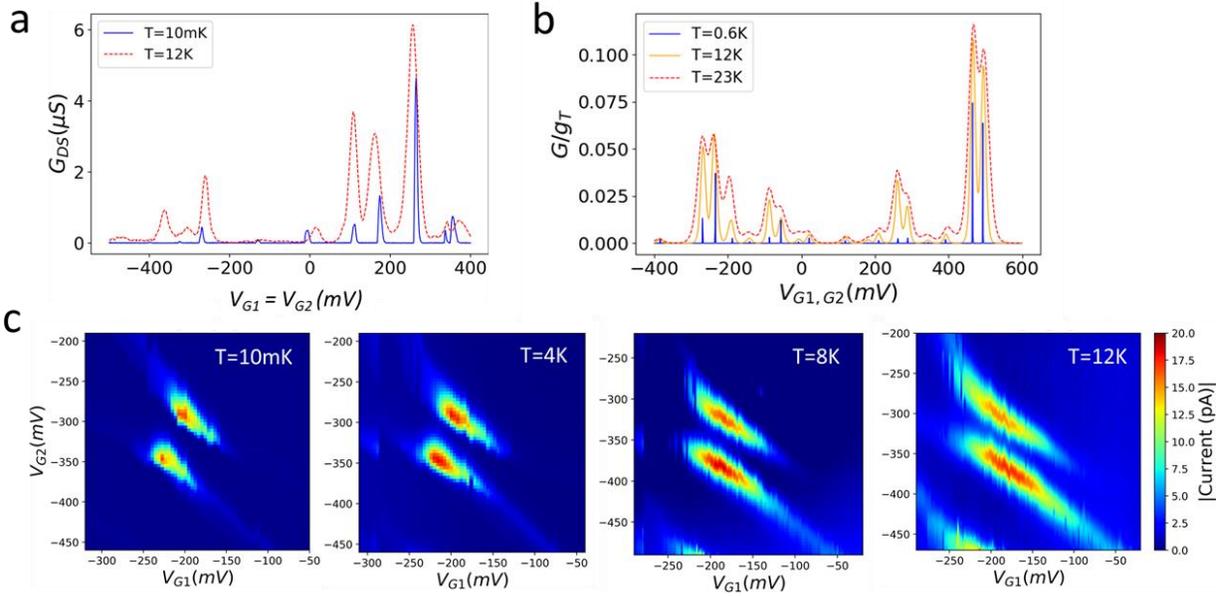

Figure 7. Characterizing thermal activation in the second array at elevated temperatures. (a) Direct current conductance measured along the dashed line in Fig. 4b at a bias $V_{bias} = 3mV$ and at temperatures $T = 10mK, 12K$. (b) Simulated Coulomb oscillation peaks in conductance of the second array at different temperatures. (See Methods) (c) Conductance at an avoided crossing region of the second array (highlighted by the circle in Fig. 4b) that is measured with a bias $V_{bias} = 6mV$ and at varying temperatures $T = 10mK, 4K, 8K, 12K$. All 4 plots in (c) share the same color bar on the right.

**Thermally Activated Transport within an Array**

Transport through the second array is of special interest because its lattice site density corresponds to the critical doping density in silicon between tunneling and band transport regimes, where transitions of the electron system from a frozen Wigner-like phase to a Fermi glass has been previously observed by increasing the temperature in a few-dopant silicon transistor.[11] Here we explore the thermal activation in the second array by monitoring the temperature evolution of transport properties through the array. The experimental temperature range investigated is limited by the temperature (~16K) at which leakage current through the Si substrate becomes comparable to conductance through the array. At elevated temperatures, the charge occupation is thermally broadened not only in the Fermi distributions within the source/drain leads, but also in the occupation of the many-body states within the array. In addition to reducing the opacity at the source and drain tunnel junctions, both thermal effects increase the ability for addition electrons to access many-body excited states, and therefore, new hopping terms within the array that were previously not accessible at base temperature. Fig. 7a plots the Coulomb oscillation in source-drain conductance at T=10mK and 12K taken along the dashed line cut in Fig. 4b using a small bias voltage of $3mV$. In this system, thermal broadening of Coulomb oscillation peaks at T=12K becomes large enough to eliminates smaller Coulomb blockade gaps within the upper and lower Hubbard bands but not the large Mott gap in between. Such thermally activated Hubbard band formation behavior within a 2D array is a key distinction from the thermal broadening in a zero-dimentional system where Hubbard bands do not exist. Due to the removal of the Coulomb blockade within each of the Hubbard bands at higher temperatures, charge stability domains characterizing fixed numbers of electrons in the array become ill-defined except for at the large Mott gap. Similar transition in transport from the collective Coulomb blockade regime to the two Hubbard bands regime is also observed in the simulated Coulomb oscillations at elevated temperatures (Fig. 7b). Fig. 7c shows the temperature evolution in conductance at an avoided crossing region of the second array (circled in Fig. 4b) at T=10mK, 4K, 8K, and 12K. Conductance at the two triple-points at the avoided crossing stretches from well-defined bias triangles to conductance lines running in parallel, in analogy to the development of crescents at the triple-points of a doube-dot system when increasing the tunnel-coupling between the dots.[39] Within the 2D array, this signature can be attributed to the additional hopping channels that become accessible at elevated temperatures (thermally activated hopping), that not only enhances conductance through the array, but also causes many-body states to become more delocalized within the array.

## Discussion

We have simulated an extended Fermi-Hubbard Hamiltonian of a 3x3 2D lattice using a series of STM-fabricated 3x3 arrays of few-dopant quantum dots and probed the many-body properties within the arrays using quantum transport measurements. By introducing two in-plane gates on either side of the array, we demonstrate gate tuning of the electron ensemble within the array as well as the flexibility of tilting the chemical potential landscape for characterizing the charge distributions of many-body ground states and resonant properties between them. By varying the lattice constants of a series of array devices at the fabrication stage, we demonstrate effective tuning of the hopping amplitude and long-range interactions within the simulated Hamiltonian. Through comparisons with theory, we can identify a finite-size analogue to a transition from Mott insulating to metallic behavior in charge distributions within the array when the lattice constant is reduced from $10.7 \pm 0.2 nm$ to $4.1 \pm 0.3 nm$. By increasing the temperature and monitoring the thermally activated Hubbard band formation near a critical lattice site density, we observe the formation of thermally activated Hubbard bands and enhanced electron transport through the array via thermally activated occupation.

These experiments confirm the viability of simulating a Fermi-Hubbard model using dopant-based 2D arrays that account for limited nonuniformity and pave the way towards more complex and accurate analog quantum simulations of extended Fermi-Hubbard Hamiltonians using dopant atoms. Tunability within the 2D array demonstrates the expected control of Coulomb interactions, hopping, filling factors, and temperature; this provides a pathway to explore previously inaccessible regions in the multi-dimensional condensed matter phase space. Extending this work to larger dopant-based arrays should allow the study of many-body localization and competing magnetic and charge order on 2D lattices, including frustrated geometries. In contrast to other quantum simulation platforms of Fermi-Hubbard Hamiltonians such as optical lattices, dopant-based artificial lattices are unique in reaching low effective temperatures and easy access to quantum transport. Furthermore, relative to gate-defined quantum dot arrays, a dopant-based system has the unique advantage in having the naturally occurring nucleus at each dopant-based lattice site, allowing simulations that include nuclear spins and hyperfine interactions inherent in real-world condensed matter physics. With continued advances to reduce disorder and improve atomic-scale precision in single-dopant placement, this method can be generalized to larger lattice arrays and other types of dopant species, such as boron[40] and arsenic,[17] potentially embedded in alternative substrate lattice environments, such as a Ge substrate.[41] The results demonstrated in this study serve as a launching point for a new class of engineered artificial lattice systems to simulate strongly correlated many-body systems in the solid states and explore less well-understood new phenomena such as the pseudogap, strange metals, topological phases, and superconductivity in the Fermi-Hubbard model and magnetic ordering and frustration in spin systems. We point out that the problem of solving the Hubbard Hamiltonian for generic parameters quickly becomes intractable with state-of-art classical computers (even for ground state properties) for systems larger than 5x5 lattice sites. Even the 3x3 arrays in this study are on the cusp of what can be numerically done today exactly for finite-temperature transport properties. We anticipate that dopant-based analog quantum simulation experiments using larger arrays should soon be able to provide answers not available to numerical calculations.

## Methods

### Device fabrication and measurements

The 3x3 few-dopant quantum dot arrays are fabricated in an ultrahigh vacuum (UHV) environment with a base pressure below $4 \times 10^{-9}$ Pa ($3 \times 10^{-11}$ Torr) using an STM tip to create atomic-scale lithographic patterns of the device by removing individual hydrogen atoms on an hydrogen-terminated, Si(100) 2x1 reconstructed surface.[42] Detailed sample preparation, UHV sample cleaning, hydrogen-resist formation, and STM tip fabrication and cleaning procedures have been published elsewhere. [43,44] The substrate is then saturation-dosed with $PH_3$ gas at room temperature, where $PH_3$ molecules selectively absorb only onto the lithographic regions where chemically reactive Si dangling bonds are exposed. A subsequent rapid thermal anneal process at 350°C for 1 min incorporates the absorbed phosphorus atoms into the Si surface lattice sites while preserving the hydrogen resist to confine dopants within the patterned regions. We embed the incorporated dopant atoms in a single crystalline silicon environment using low-temperature Si epitaxial overgrowth with an optimized locking layer to suppress dopant movement at the atomic scale during epitaxial overgrowth [25,44] Finally, ohmic contacts to the buried device are formed using a low thermal budget contacting technique.[26] The number of incorporated P atoms in each quantum dot can be estimated by analyzing the lithographic sites that are available for P incorporation.[30] See Supplementary Materials Section S2 for analyzing the best estimates of dopant numbers at each lattice site. Using the Si(100) 2x1 surface reconstruction lattice as an atomically precise ruler, the lattice constant of the fabricated arrays (center-to-center distance between adjacent quantum dots) can be precisely characterized. The STM-patterned in-plane source/drain leads and gates are saturation-doped[45] with a dopant density of $\sim 2 \times 10^{14}/cm^2$, corresponding to a 3D doping density of $\sim 2 \times 10^{21}/cm^3$ that is approximately three orders of magnitude above the metal-insulator transition, allowing quasi-metallic conduction in all electrodes. We carry out direct-current (DC) transport measurement of the device inside a dilution refrigerator at a base temperature of ~10mK.

### Numerical simulation

We have constructed a time-independent extended Hubbard model to simulate the 3x3 quantum dot array system following the analysis first introduced by Das Sarma and coworkers[46] for semiconductor quantum dots and more recently by Ginossar and coworkers[27] for dopant arrays in silicon. In our extended Hubbard model (see Eq. 1), we include one spinful orbital per dot near the Fermi level, which means a maximum of 18 valence electrons can be added onto the array. We include only nearest-neighbor hopping. We formulate the electrostatic potential landscape as well as on-site and long-range electron-electron Coulomb interactions using capacitance matrix calculations based on the STM-lithography patterns of the device geometry, where a lateral seam of 2.5 nm and a vertical thickness of 2 nm have been included to account for the Bohr-radius-like electron density extension beyond the lithographic patterns, shown previously to be necessary to reproduce the experimental capacitance values in STM-patterned Si:P devices.[15,47] We estimate the number of dopant atoms in the quantum dots by characterizing the binding energies of similar few-dopant quantum dots with a similar lithographic patch (See Supplementary Materials Section S3). We approximate the screened on-site electron-ion Coulomb interaction in a quantum dot using the charge-neutral binding energy of the estimated dopant cluster.[28] We adopt a Fermi level of -80meV with respect to Si conduction band edge in the saturation doped, in-plane source/drain, and gate electrodes.[48,49] We approximate the long-range electron-ion Coulomb attractions using a point charge approximation.[27] More details about model parameter extractions are provided in the Supplementary Materials. Exchange interactions and longer-range hopping terms are not implemented in our model. We assume the lead's tunnel coupling to the dot array is sufficiently weak so that the dot array can be treated as an isolated system when solving for the eigenstates.

At each gate-gate point, we use exact diagonalization to solve for the ground and excited many-body eigenstates of the Hamiltonian, whose matrix is represented in the Fock basis, for particle numbers ranging from 0 to 18 in the 3x3 array system. The charge stability diagram simulations for the first and second array utilize the classical capacitance coupling and best-estimated numbers of dopants per site based on the STM-patterned device geometries (See Supplementary Materials Section S2). The hopping amplitudes are set as t=0.5 meV, t=2 meV and t=8 meV for simulating the first, second, and third arrays, respectively, based on the designed lattice constants in the arrays. Charge stability diagrams and ground state charge distributions are obtained using the Lanczos algorithm after further block-diagonalizing the Hamiltonian matrices based on the total spin in the z direction. The numerically simulated conductance results in the main text are linear response conductance through the array at finite temperatures and zero bias that is calculated following the formalism of Ginossar et al.[24], which uses Fermi's golden rule for the tunneling rate to/from the leads and a generalization of the transport equations developed by Beenakker [50] for a single quantum dot. For that, we implement full re-orthogonalization of vectors in the Krylov space in our Lanczos routine after each iteration and seek convergence for 25 to ~40 low-lying states if the size of the Hilbert space for a particular spin and particle number sector is greater than 2000. For smaller sizes, we perform full diagonalization using Intel's math kernel library.[51]

Adopting a similar notation as Ginossar et al.[24], we write the conductance as

$$G = g_T \sum_{n,m} \sum_{\alpha,\beta} \frac{M^{(L),n,m}_{\alpha,\beta} M^{(R),n,m}_{\alpha,\beta}}{M^{(L),n,m}_{\alpha,\beta} + M^{(R),n,m}_{\alpha,\beta}} \times P^{n,m}_\alpha \left[1 - f_{FD}\left(E^{n,m}_\alpha - E^{n-1,m-1}_\beta\right)\right]$$

Where $g_T = e^2 \Gamma/(\hbar\, kT)$,

$$P^{n,m}_\alpha = \frac{exp[-(1/kT)E^{n,m}_\alpha]}{\sum_{n,m,\alpha} exp[-(kT)E^{n,m}_\alpha]}$$

and

$$M^{(L)n,m}_{\alpha,\beta} = \sum_{j \in cL} \left|\langle \Psi^{n,m}_\alpha | c^\dagger_{j\uparrow} | \Psi^{n-1,m-1}_\beta \rangle\right|^2$$

where the last sum runs over sites on the left edge of the array, closest to the source lead. $M_{\alpha,\beta}^{(R)n,m}$ has a similar formula with a sum that runs over sites on the right edge of the array. $E_\alpha^{n,m}$ is the $\alpha^{th}$ eigenenergy of the many-body wavefunction, $\Psi_\alpha^{n,m}$ with $n$ particles and a total spin of $m$. $k$ is Boltzmann's constant, and $f_{FD}$ is the Fermi-Dirac distribution function. $\Gamma$ is the hopping amplitude for electrons to/from the leads. In out plots, we show results for the dimensionless quantity $G/g_T$. We use spin inversion symmetry to simplify our calculations by considering the transport of only spin-up electrons through the array and inserting an overall factor of two in $G$.

## Acknowledgements

We acknowledge helpful discussions with Emily Townsend, Neil Zimmerman, Alexey Gorshkov, and Maicol A. Ochoa. This research was funded in part by the Department of Energy Advanced Manufacturing Office Award Number DE-EE0008311 and by a National Institute of Standards and Technology (NIST) Innovations in Measurement Science award, "Atom-Based Devices: Single Atom Transistors to Solid State Quantum Computing." E. K. was supported by the National Science Foundation under Grant No. DMR- 1918572. This work was performed in part at the Center for Nanoscale Science and Technology NanoFab at NIST.

## Contributions

X.W. performed the experiments, X.W. and R.S. analyzed the data, . X. W., J. W., and P. N. fabricated the devices. X. W. and E. K. formulated the theoretical model and carried out the numerical simulations. X. W., R. S., E. K,  F. F., and G. B. contributed to the theoretical interpretation of the data. X. W., A. R., R. K., F. F., and R. S. contributed to electrical measurements. X.W., E. K., and R. S. wrote the manuscript with comments from all co-authors. R.S. conceived and supervised the project.

## Supplementary Materials

### S1. Capacitance Modeling of the 3x3 Array Devices

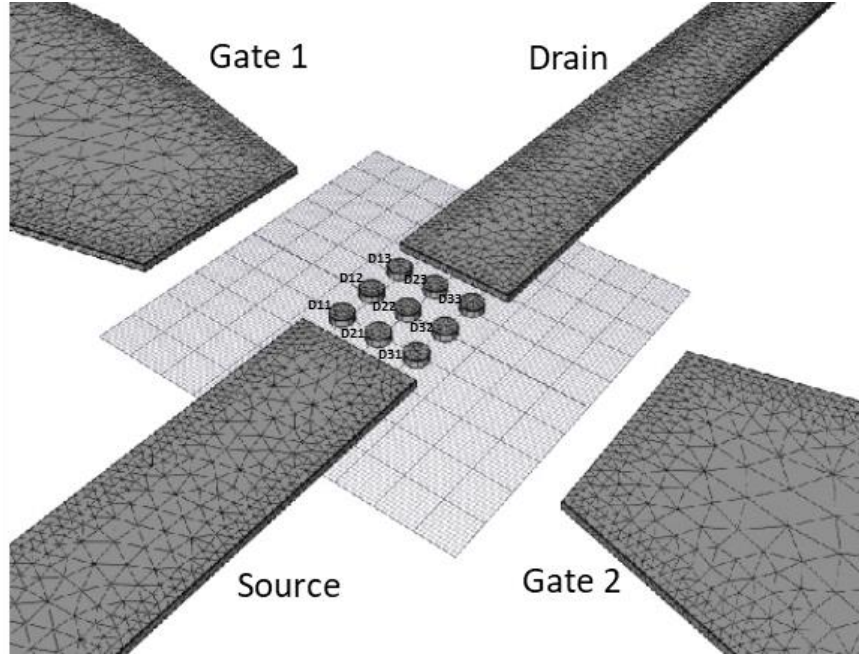

Figure S1. Classical capacitance model of the 3x3 array buried in the device geometry. The figure shows a perspective view of the mesh geometry of the first array for modeling the capacitance matrix of the device.

Gate-tunability of the chemical potential landscape within the array is achieved via classical capacitive coupling between the two in-plane gates and the lattice site in the array. From the equivalent circuit model under the constant interaction approximation,[53] there exists constant capacitive coupling between the electrons at a lattice site i and a metallic electrode $l$, denoted as $C_{l,i}$, and between electrons at different lattice sites, denoted as $C_{i,j}$. Here $l \in [D, S, G1, G2]$ for the drain, source, gate1, and gate2 electrodes, and $i, j \in [11,12,13,21,22,23,31,32,33]$ for the quantum dot sites from $D11$ to $D33$. The total capacitance of site i is $C_{\Sigma i} = \sum_l C_{l,i} + \sum_{j \neq i} C_{i,j}$. Therefore, the capacitance matrix of the array system can be expressed as,

$$\hat{C} = \begin{bmatrix} C_{\Sigma 11} & -C_{11,12} & -C_{11,13} & -C_{11,21} & -C_{11,22} & -C_{11,23} & -C_{11,31} & -C_{11,32} & -C_{11,33} \\ -C_{11,12} & C_{\Sigma 12} & -C_{12,13} & -C_{12,21} & -C_{12,22} & -C_{12,23} & -C_{12,31} & -C_{12,32} & -C_{12,33} \\ -C_{11,13} & -C_{12,13} & C_{\Sigma 13} & -C_{13,21} & -C_{13,22} & -C_{13,23} & -C_{13,31} & -C_{13,32} & -C_{13,33} \\ -C_{11,21} & -C_{12,21} & -C_{13,21} & C_{\Sigma 21} & -C_{21,22} & -C_{21,23} & -C_{21,31} & -C_{21,32} & -C_{21,33} \\ -C_{11,22} & -C_{12,22} & -C_{13,22} & -C_{21,22} & C_{\Sigma 22} & -C_{22,23} & -C_{22,31} & -C_{22,32} & -C_{22,33} \\ -C_{11,23} & -C_{12,23} & -C_{13,23} & -C_{21,23} & -C_{22,23} & C_{\Sigma 23} & -C_{23,31} & -C_{23,32} & -C_{23,33} \\ -C_{11,31} & -C_{12,31} & -C_{13,31} & -C_{21,31} & -C_{22,31} & -C_{23,31} & C_{\Sigma 31} & -C_{31,32} & -C_{31,33} \\ -C_{11,32} & -C_{12,32} & -C_{13,32} & -C_{21,32} & -C_{22,32} & -C_{23,32} & -C_{31,32} & C_{\Sigma 32} & -C_{32,33} \\ -C_{11,33} & -C_{12,33} & -C_{13,33} & -C_{21,33} & -C_{22,33} & -C_{23,33} & -C_{31,33} & -C_{32,33} & C_{\Sigma 33} \end{bmatrix}$$

The electrostatic charge at site i can be expressed as $q_i = -n_i e + \sum_l C_{l,i} V_l$, where $n_i \in [0,1,2]$ is the integer number of electrons added onto the site via hopping, and $C_{l,i} V_l$ is the capacitively induced charge when applying a voltage $V_l$ on electrode $l$. Here, e is the absolute charge value of an electron;

and we neglect the background induced charges from the substrate. The classical charge at the array can be expressed as,

$$\hat{Q} = \begin{bmatrix} q_{11} \\ q_{12} \\ q_{13} \\ q_{21} \\ q_{22} \\ q_{23} \\ q_{31} \\ q_{32} \\ q_{33} \end{bmatrix}$$

The electrostatic potential $\hat{V}$ and the total electrostatic energy P of the array can be obtained from,

$$\hat{V} = \frac{\hat{Q}}{\hat{C}}$$

$$P = \frac{1}{2}\hat{V} \cdot \hat{Q}$$

Here $P$ is a function of integer numbers of electrons at each site $n_i$ and the voltage values at each electrode $V_l$. When $n_i = 0$ for all sites, we denote the total eledctrostatic energy as $P_0$. At a given set of voltage conditions, we denote $p_i$ and $U_i$ as the electrostatic chemical potential and on-site charging energy to add the first and second electrons, respectively, onto site $i$, which can be numerically calculated from,

$$p_i = P(0, \ldots, n_i = 1, \ldots, 0) - P_0$$

$$U_i = [P(0, \ldots, n_i = 2, \ldots, 0) - P(0, \ldots, n_i = 1, \ldots, 0)] - p_i$$

The electrostatic mutual charging energy $U_{i,j}$ between site $i$ and site $j$ can be calculated from,

$$U_{i,j} = [P(0, \ldots, n_i = 1, \ldots, 0, \ldots, n_j = 1, \ldots, 0) - P_0] - p_i - p_j$$

Note, it can be numerically verified that $U_i$ and $U_{i,j}$ are independent of $V_l$. On the other hand, $U_i$ and $U_{i,j}$ depend critically on the size of the metallic disk that is used to represent a quantum dot in the classical capacitance model. In each of the array devices in this study, we chose the disk size so that the average value of $U_i$ (see Table S1(a)) matches (within +/-2meV range) the average expected on-site charging energy values of few-dopant clusters from previous atomic tight-binding calculations (See the last column in Table S3).

We denote $\alpha_{l,i}$ as the lever arm of an electrode l to a lattice site i, which is defined as the ratio between $\Delta p_{l,i}$, the shift in the electrostatic chemical potential at site i in response to $\Delta V_l \times e$, the change in the electrostatic potential at electrode l.

$$\alpha_{l,i} = \frac{\Delta p_{l,i}}{\Delta V_l \times e}$$

As described in the main text, the gate-gate slope at a charge addition boundary corresponds to the gate-gate level arm ratio of the added electron, that is, the conducting electron, and characterizes whether the single electron is added onto sites in the upper, middle, or lower row in the array. Here we define the effective level-arm of gate $l$ to an $(N+1)th$ addition-electron, which, with finite hopping, has the charge distribution $\sum_{i,\sigma} \Delta n_{i,\sigma}$, as,

$$\alpha_l(N \to N+1) = \sum_i \Delta n_i \cdot \alpha_{l,i}$$

Here $\Delta n_i$ represents the change in occupation at site $i$ when a single electron is added onto the array, altering the charge distribution of the array from an eigenstate of a total charge number $N$ to that of an eigenstate of $N+1$. The charge number conservation condition requires that $\sum_i \Delta n_i = 1$. On a numerically simulated charge stability diagram, we can extract the added electron occupations at charge addition boundaries, see red-blue binary plots in Fig. 3a in the main text, for instance, and calculate the lever arm ratio $\frac{\alpha_{G1,N\to N+1}}{\alpha_{G2,N\to N+1}}$ at each pixel point. In the right panels in Figs. 5d and 5e in the main text, we plot the histogram distributions of the lever arm ratio calculated from the simulated charge stability diagrams in Fig. 2d for the first array and in Fig. 5a for the second array, respectively.

On a conductance map spanned by voltages of the two in-plane gates, the gate-gate lever arm ratio equals the slope of the conduction line $\frac{\alpha_{G1}}{\alpha_{G2}} = \frac{-\Delta V_{G2}}{\Delta V_{G1}}$. Indeed, we observe qualitative agreement between the measured slope histograms and calculated lever arm ratio histograms in Figs. 5d and 5e.

(a)

| On-site electron-electron Coulomb repulsion $U_i$ (meV) | | | |
|---|---|---|---|
| | First Array | Second Array | Third Array |
| $U_{11}$ | 45.57 | 46.95 | 46.81 |
| $U_{12}$ | 47.95 | 47.89 | 45.81 |
| $U_{13}$ | 45.57 | 46.95 | 46.93 |
| $U_{21}$ | 44.66 | 45.61 | 44.14 |
| $U_{22}$ | 47.33 | 46.32 | 42.97 |
| $U_{23}$ | 44.66 | 45.61 | 44.15 |
| $U_{31}$ | 45.57 | 46.95 | 46.26 |
| $U_{32}$ | 47.95 | 47.89 | 45.23 |
| $U_{33}$ | 45.57 | 46.95 | 46.00 |

(b)

| Long-range electron-electron Coulomb repulsion $U_{i,j}$ (meV) | | | |
|---|---|---|---|
| | First Array | Second Array | Third Array |
| $U_{11,12}$ | 8.19 | 13.78 | 20.25 |
| $U_{11,13}$ | 2.89 | 5.94 | 10.92 |
| $U_{11,21}$ | 6.72 | 13.77 | 19.06 |
| $U_{11,22}$ | 4.82 | 9.01 | 13.94 |
| $U_{11,23}$ | 2.14 | 4.80 | 9.20 |
| $U_{11,31}$ | 2.30 | 6.05 | 10.08 |
| $U_{11,32}$ | 2.42 | 5.45 | 9.10 |
| $U_{11,33}$ | 1.36 | 3.47 | 6.94 |
| $U_{12,13}$ | 8.19 | 13.78 | 20.27 |
| $U_{12,21}$ | 4.77 | 9.02 | 13.96 |
| $U_{12,22}$ | 8.90 | 15.00 | 19.06 |
| $U_{12,23}$ | 4.77 | 9.02 | 14.02 |
| $U_{12,31}$ | 2.42 | 5.45 | 9.08 |
| $U_{12,32}$ | 3.50 | 7.16 | 10.71 |
| $U_{12,33}$ | 2.42 | 5.45 | 9.10 |
| $U_{13,21}$ | 2.14 | 4.80 | 9.09 |
| $U_{13,22}$ | 4.82 | 9.01 | 13.88 |
| $U_{13,23}$ | 6.72 | 13.77 | 19.15 |
| $U_{13,31}$ | 1.36 | 3.47 | 6.89 |
| $U_{13,32}$ | 2.42 | 5.45 | 9.07 |
| $U_{13,33}$ | 2.30 | 6.05 | 10.10 |
| $U_{21,22}$ | 7.89 | 13.05 | 18.68 |
| $U_{21,23}$ | 2.60 | 5.66 | 10.35 |
| $U_{21,31}$ | 6.72 | 13.77 | 18.86 |
| $U_{21,32}$ | 4.77 | 9.02 | 13.79 |
| $U_{21,33}$ | 2.14 | 4.80 | 9.04 |
| $U_{22,23}$ | 7.89 | 13.05 | 18.67 |
| $U_{22,31}$ | 4.82 | 9.01 | 13.74 |
| $U_{22,32}$ | 8.90 | 15.00 | 18.85 |
| $U_{22,33}$ | 4.82 | 9.01 | 13.73 |
| $U_{23,31}$ | 2.14 | 4.80 | 8.98 |
| $U_{23,32}$ | 4.77 | 9.02 | 13.78 |
| $U_{23,33}$ | 6.72 | 13.77 | 18.85 |
| $U_{31,32}$ | 8.19 | 13.78 | 19.79 |
| $U_{31,33}$ | 2.89 | 5.94 | 10.65 |
| $U_{32,33}$ | 8.19 | 13.78 | 19.72 |

(c)

| | Gates' lever arms to each quantum dot $\alpha_{l,i}$ | | | | | | | | |
|---|---|---|---|---|---|---|---|---|---|
| | First Array | | | Second Array | | | Third Array | | |
| | $\alpha_{G1,i}$ | $\alpha_{G2,i}$ | $-\alpha_{G1,i}/\alpha_{G2,i}$ | $\alpha_{G1,i}$ | $\alpha_{G2,i}$ | $-\alpha_{G1,i}/\alpha_{G2,i}$ | $\alpha_{G1,i}$ | $\alpha_{G2,i}$ | $-\alpha_{G1,i}/\alpha_{G2,i}$ |
| Dot11 | 0.153 | 0.084 | -1.834 | 0.158 | 0.101 | -1.563 | 0.096 | 0.093 | -1.042 |
| Dot12 | 0.186 | 0.107 | -1.746 | 0.175 | 0.116 | -1.514 | 0.103 | 0.099 | -1.043 |
| Dot13 | 0.153 | 0.084 | -1.834 | 0.158 | 0.101 | -1.563 | 0.097 | 0.093 | -1.041 |
| Dot21 | 0.105 | 0.105 | -1.000 | 0.121 | 0.121 | -1.000 | 0.082 | 0.102 | -0.806 |
| Dot22 | 0.136 | 0.136 | -1.000 | 0.138 | 0.138 | -1.000 | 0.087 | 0.109 | -0.799 |
| Dot23 | 0.105 | 0.105 | -1.000 | 0.121 | 0.121 | -1.000 | 0.082 | 0.102 | -0.802 |
| Dot31 | 0.084 | 0.153 | -0.545 | 0.101 | 0.158 | -0.640 | 0.073 | 0.119 | -0.610 |

| | | | | | | | | |
|---|---|---|---|---|---|---|---|---|
| Dot32 | 0.107 | 0.186 | -0.573 | 0.116 | 0.175 | -0.661 | 0.077 | 0.128 | -0.605 |
| Dot33 | 0.084 | 0.153 | -0.545 | 0.101 | 0.158 | -0.640 | 0.073 | 0.121 | -0.600 |

Table S1. Calculated on-site electron-electron interactions ($U_i$), long-range electron-electron interactions ($U_{i,j}$), and gate lever arms ($\alpha_{l,i}$) in the three arrays presented in this study.

## S2. Image Analysis of STM-patterned Quantum Dots

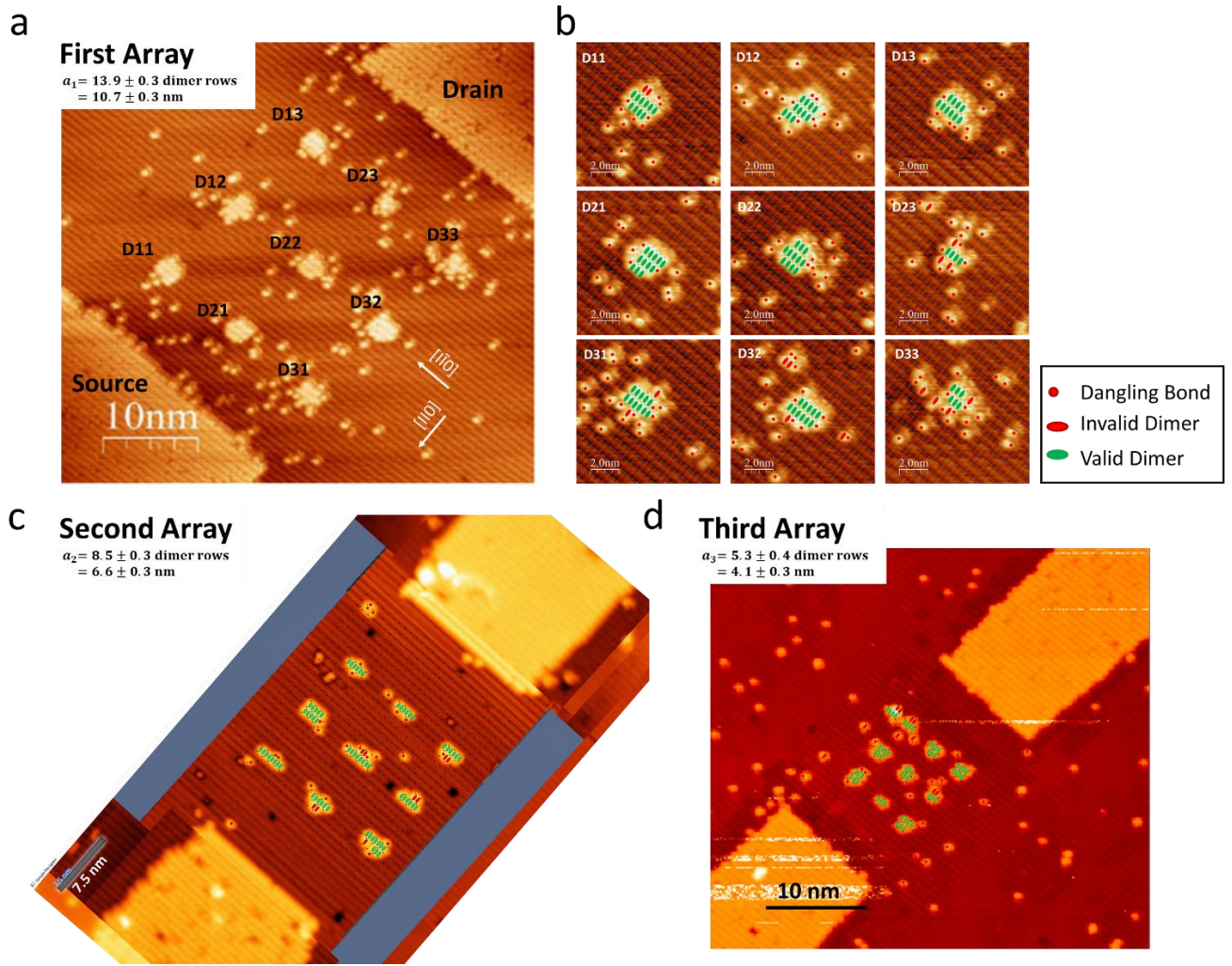

Figure S2. Estimating the number of incorporated dopant atoms at each quantum dot based on atomic-resolution STM images of hydrogen lithography patterns before phosphine dosing. Previous studies[13,30,31] have shown that the number of incorporated P atoms at STM-patterned quantum dot can be estimated by counting the number of hydrogen-desorbed dimers that are valid for P incorporation. We have carried out previous studies to verify the dopant atom incorporation probability in patterned devices by comparing the STM lithography to few atom quantum dot spectroscopy as described in more detail in Supplementary Materials Section S3. Having at least three adjacent H-desorbed dimers within the same dimer row is a necessary condition to incorporate one P dopant into the surface silicon

lattice. (a) Atomic resolution STM image of the central region of the first array device taken after hydrogen lithography but before phosphine dosing. (b) Zoom-in images at each quantum dot in (a) with overlayed grids of Si(100) 2x1 surface reconstruction unit cells and identifiers of desorbed dangling bonds and dimers at each array site. (c) (d) Atomic resolution STM image of the central region of the second array device (c) and the third array device (d) after hydrogen lithography but before phosphine dosing, overlayed with surface lattice grids and identifiers of desorbed dangling bonds and dimers at each array site.

At each quantum dot in the STM images of the arrayed device hydrogen lithography patterns (see Figure S2), we identify single dangling bonds, valid and invalid dimer sites for P incorporation sites using red dots, green ellipses, and red ellipses, respectively. We define the best estimate of the number of incorporated P atoms at each site by requiring three contiguous dimers which are then counted as capable of incorporating a single P dopant. The upper bound of our estimate of the number of incorporated P atoms is set by 25% of the number of dangling bonds sites within allowed dimers, according to the 0.25 monolayer of P incorporation density in saturate-doped Si:P $\delta$-layers.[54] To estimate the lower bound of incorporated P atoms we follow Fuchsle et al.,[31] where it is found that, with a saturation dose of phosphine gas at room temperature, the P incorporation density in nm scale desorbed areas decreases to 0.09 monolayer of contiguous desorbed areas within the quantum dot pattern due to competition for dangling bond sites to lose H atoms from absorbed $PH_x$ (x=1, 2) during incorporation. These estimates agree with our independent evaluation of incorporation probabilities using few dopant atom transistors as described in S3. We round the estimated lower and upper bound dopant numbers to their nearest integers. Table S1 lists the best estimated numbers of dopants at each quantum dot and the lower and upper bounds of the estimates. While it is possible to count the number of incorporated dopants directly by imaging the quantum dots after P incorporation using STM, we avoid this time-consuming step to minimize the exposure of the reactive patterned device to contamination before epitaxial encapsulation as well as the risk of unintentional tip-surface interactions that could introduce atom-scale contaminations and defects.

| Site Index | Estimated Number of Dopants | | | | | | | | |
|---|---|---|---|---|---|---|---|---|---|
| | First Array | | | Second Array | | | Third Array | | |
| | Lower Bound of Estimate | Best Estimate | Upper Bound of Estimate | Lower Bound of Estimate | Best Estimate | Upper Bound of Estimate | Lower Bound of Estimate | Best Estimate | Upper Bound of Estimate |
| D11 | 2 | 3 | 5 | 1 | 3 | 4 | 1 | 2 | 3 |
| D12 | 2 | 3 | 5 | 2 | 3 | 5 | 1 | 2 | 3 |
| D13 | 2 | 3 | 5 | 1 | 2 | 3 | 1 | 2 | 4 |
| D21 | 2 | 3 | 6 | 1 | 2 | 3 | 0 | 1 | 1 |
| D22 | 2 | 3 | 6 | 1 | 3 | 4 | 1 | 2 | 3 |
| D23 | 0 | 1 | 2 | 1 | 2 | 3 | 1 | 2 | 3 |
| D31 | 2 | 3 | 5 | 2 | 3 | 5 | 1 | 2 | 3 |
| D32 | 2 | 3 | 5 | 1 | 2 | 3 | 0 | 1 | 1 |
| D33 | 1 | 2 | 3 | 1 | 2 | 3 | 1 | 2 | 4 |

Table S2. Estimated number of incorporated P atoms at each quantum dot. The ranges of estimates are based on analysis in Fig. S2. See descriptions in Supplementary Materials Section S2 for the criteria of the lower bound, the upper bound, and the optimum of the estimates in this table.

| Dopant Number | $E_b$ | $U_i$ |
|---|---|---|
| 1P | $-47$ | 44 |
| 2P | $-70 \pm 10$ | $45 \pm 10$ |
| 3P | $-81 \pm 6$ | $46 \pm 7$ |

Table S3. Binding energy $E_b$ and addition energy $U_i$ (for the relevant D⁰ to D⁻ transitions near the Fermi level) of few-P cluster quantum dots. $E_b$ values are with respect to the conduction band edge. The variations for 2P and 3P clusters represent one sigma (70%) in cluster configuration distributions.[28] We account for the variations in the best-estimate numbers of dopant atoms per site (Table S2) by adopting the corresponding binding energy (average value) as listed in the second column. The charging energy values in the last column are used to choose the dot size in the capacitance model so that the calculated on-site charging energy (See Table S1) matches (within a few meV) to the atomistically calculated charging enery values.

**S3. Characterizing an individual few-dopant quantum dot**

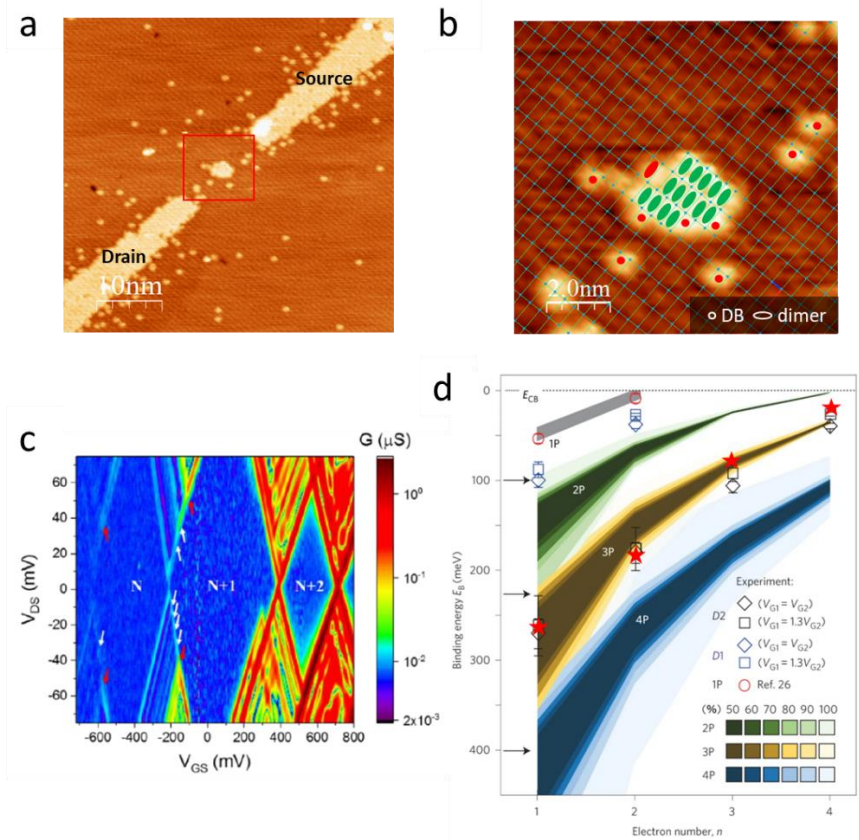

Figure S3. An example of few-dopant quantum dot spectroscopy. (a) STM images of the central region of the few-dopant quantum dot after hydrogen lithography, but before PH$_3$ dosing. (b) Close-up STM image of the H-desorbed quantum dot region. The number of exposed Si dangling bonds (DB) and dimers can be counted by overlaying the Si (100) 2x1 surface reconstruction lattice grids with the STM images after hydrogen lithography. The allowed and forbidden P incorporation sites are highlighted in green and red respectively. (c) the low-temperature (T=4 K) differential conductance charge stability diagram. As indicated by arrows at the $N \leftrightarrow N+1$ and $N-1 \leftrightarrow N$ transitions, there appears symmetric resonant tunneling features at positive and negative biases, indicating approximately equal tunnel coupling between the dot and the drain and source leads. The occupation number of the dot is expressed using an integer $N$. (d) Following the method as described by Weber and co-workers,[28] we extract the binding energy spectrum of the few-dopant quantum dot from (c) and overlay the extracted binding energy levels (red stars) on the experimental and theoretical binding energy spectrum previously published by Weber and co-workers.[28] The overlay indicates that there are 3 dopant atoms incorporated in the dot shown in (b).

## S4. Impact of Hopping, Long-Range Interactions, and Disorder on the Addition Energy Spectrum

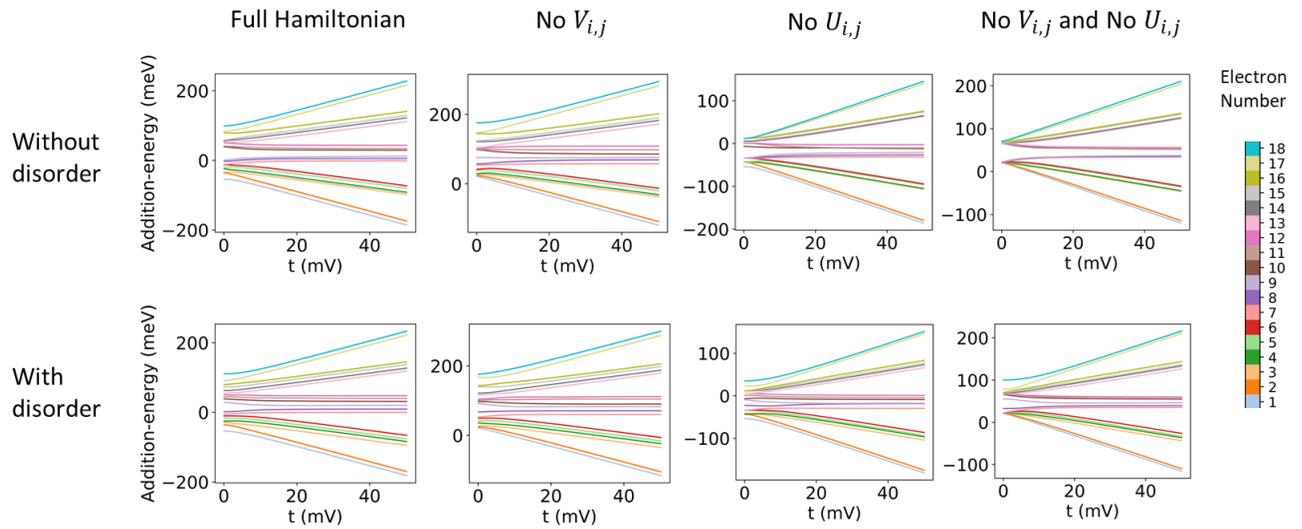

Figure S4. Impact of long-range electron-electron Coulomb repulsion ($U_{i,j}$), long-range electron-ion core Coulomb attraction ($V_{i,j}$), and hopping amplitude on the charge addition energy spectrum. The panels plot the simulated charge addition energy spectrum of the first array as a function of hopping amplitude with all leads and gate electrodes at zero ground potential. The upper panels assume identical three-dopant quantum dots while the lower panels take into account the estimated disorder in the array (see Tables S1 and S2). We have calculated with different sets of disorder configuratios and found that they do not alter the qualitative features as shown in the plots in the lower panels here. Panels in the first column are calculated using the full Hamiltonian (Equations 1 in the main text). Long-range electron-ion core Coulomb attraction and/or long range electron-electron Coulomb repulsion are turned off in subsequent panels. The 18 addition energy levels correspond to a total of 18 excess electrons that the model allows to be added onto the 3x3 array system.

The ground state addition energy levels for adding the $N^{th}$ electron onto the array, $E^0_{add}$ is defined as

$$E^0_{add}(N) = E^0_N - E^0_{N-1}$$

here $E^0_N$ is the N-electron ground eigen-energy of the array.

Figure S4 shows the calculated charge addition energy spectra of a 3x3 dot-array either ignoring or including long-range electron-electron and electron-ion Coulomb interactions. The upper panels in Figure S4 assume no disorder in the artificial lattice sites (identical three-dopant quantum dots) and hopping amplitude and with all lead and gate electrodes grounded. The hopping amplitude (t) is swept from the weak tunnel coupling regime where t<<U to the intermediate/strong tunnel coupling regime where t ~ U.  At t=0 and in the absence of inter-site interactions, the Mott gap separation equals the charging energy at each quantum dot. In this case, transport through the array is in the *classical Coulomb blockade* regime because the charge addition spectrum is solely determined by the on-site charging energy of each quantum dot. At t>0 and in the presence of inter-site interactions, both hopping and inter-site interactions broaden the Hubbard bands and suppress the Mott gap at half-filling. In this case, transport through the array is in the *collective Coulomb blockade* regime[7,33] because the charge addition spectrum is determined by the combined effects of the on-site and inter-site interactions within the array. Similar computational observations have been previously reported by Le, Fisher, and Ginossar[27] in dopant-based arrays of similar size.[27] In this study, we extend previous efforts in simulating dopant arrays by including two experimentally defined in-plane gates in the extended Hubbard model and explore their impact on the charge distributions and addition energy spectrum in the array.

## S5. Impact of Disorder on Charge Stability Diagrams

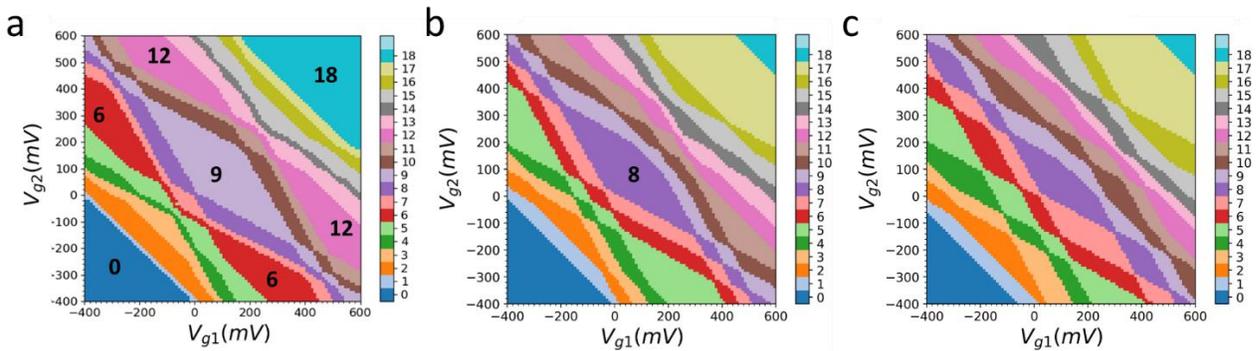

Figure S5. Understanding the impact of disorder and gate potential gradient on the ground eigenstate charge stability diagram. The simulated charge stability diagram in (a) is based on an idealized array with no disorder; all lattice sites are of identical three-dopant cluster quantum dot (binding energy $E_b = -81$meV) and all nearest-neighbor hopping amplitudes $t = 0.5$meV.  Charge occupation numbers at representative regions are highlighted in the diagram. When the two in-plane gates are tied and sweeping together along the diagonal direction, a Mott gap is observed at half-filling (occupation = 9), separating the lower Hubbard band from the upper Hubbard band. When a differential voltage is applied across the two in-plane gates and swept from zero to a large differential value, the Mott gap at half-filling closes, and two addition-energy gaps at one-third (occupation = 6) and two-thirds (occupation = 12) filling opens. This process occurs due to the lowering of the chemical potential of the first and the second rows of lattice sites causing the electrons to energetically favor doubly occupying those rows and overcoming the on-site charging energy (U), as opposed to half-filling of the entire array.In (b), we introduce disorder in the number of dopant atoms per site based on the best estimates from experimental lithographic patterns shown in Fig. 1c in the main text: one dopant at D23 ($E_b = -47$meV), two dopants at D33 ($E_b = -70$meV), and three dopants for the rest of the dots ($E_b = -81$meV). Because of the higher energy cost to add the first electron onto D23, the central Mott gap along the diagonal direction now appears with charge number 8. In (c), the dopant number at each site is the same as those in (b), however, random noise from a Gaussian distribution with zero mean and standard deviation of 10meV has been added to $E_b$ to gain a better idea about the effect of variations in the number of dopants at each site. Note that (c) is not averaged over disorder realizations; it represents a typical case in which there is significant variation in $E_b$ across the array. Along the diagonal direction, the added disorder further broadens the upper and lower Hubbard bands and suppresses the Mott gap.

## S6. Eigenstates, Charge Addition, and Resonance at Avoided Crossings

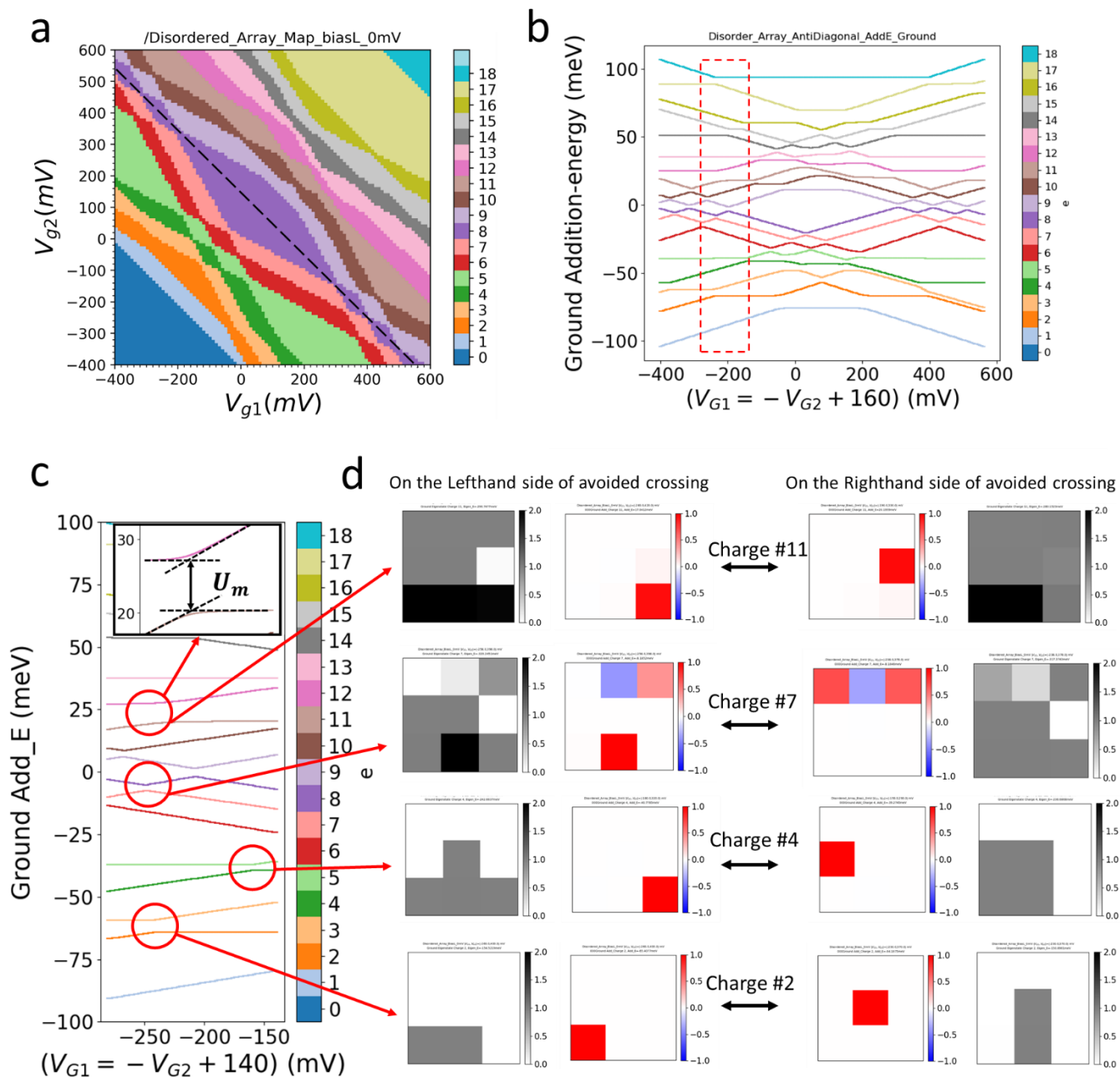

Figure S6. Simulated charge addition energy spectrum and charge distribution at avoided crossings. (a) Simulated ground eigenstate charge stability diagram of the first 3x3 array, reproduced from Fig. 2d in the main text. (b) calculated ground-state charge-addition-energy level spectrum along the dashed line in (a). See Supplementary Materials Section S4 for the definitions of the ground-state charge-addition-energy levels. The addition energy spectrum is plotted with respect to the source/drain leads' Fermi level, which sets the zero-energy level in the addition energy spectra for adding electrons from the source/drain leads. The 18 electron addition energy levels correspond to a total of 18 excess electrons that can be added to the 3x3 array at gate voltage conditions along the dashed line in (a). We emphasize the apparent similarity between the charge addition boundaries in the gate-gate charge stability map and the charge addition energy levels in the addition energy spectrum, manifesting the same charge addition response to the chemical potential gradient across the array in the gate-gate direction; the charge addition energy is flat versus the potential gradient when an electron is added to the middle row sites; the charge addition energy increases (decreases) with increased negative voltage on the upper gate when an electron is added to the upper (lower) row sites. (c) Charge occupation configurations on the left and right sides of selected avoided crossings in the addition energy spectrum region that are highlighted in (b). The charge configurations become resonant at the avoided crossing. As discussed in (b), it is energetically favorable for an electron to be added to different lattice sites on the left and right sides of an avoided crossing, as indicated by the different addition energy slopes on either side. The energy separation at an avoided crossing in the addition energy spectrm characterizes $(U_m + 2t')$, where $U_m$ is the mutual charging energy between the two involved charge addition distributions and $t'$ is the resonant tunnel coupling between the two charge distribution configurations. The subplot in (c) illustrates that the $U_m$ contribution at an avoided crossing can be obtained by extrapolating the linear sections on both sides of the avoided crossing, in analogy to standard practices in double-dot systems. Since the system is in the strong interaction regime ($U_i, U_{i,j} \gg t$), $U_m$ dominates the avoided crossing separation. (d) Schematic illustration of eigenstate charge distributions and single charge addition at the left and right sides of the avoided crossings as highlighted in (c). Taking charge addition #7 as an example, the black-white plots represent the charge distributions of many-body ground states when there are 7 electrons in the array. The red-blue plots represents the change in charge distributions when adding the 7[th] electron onto the array and the system changes from a 6-electron ground state to a 7-electron ground state. At the avoided crossings when adding charges #2, #4, and #11, the involved charge addition sites are single lattice sites, and the avoided crossing separation is large (small) when the two resonant sites are close to (far apart from) each other. At the avoided crossing for adding charge #7, due to finite hopping amplitude, the addition of a single electron leads to changes in occupation at multiple lattice sites. The resonance between the two coupled many-body charge configurations is a manifestation of the complex many-body interactions, such as the mutual charging energy (long range Coulomb interactions) and tunnel coupling, between many-body states in the array system.

| Representative Avoided-crossings in Figure S6 | Addition Energy at the Avoided-crossing (meV) | Charge-addition sites | | Spatial distance between the two charge-addition distributions | Mutual charging energy ($U_m$) between the two charge-addition distributions (meV) | 2 × Tunnel coupling ($2t'$) between the two charge-addition distributions (meV) |
|---|---|---|---|---|---|---|
| | | On the Lefthand side of avoided crossing | On the righthand side of avoided crossing | | | |
| Add Charge #2 | 4.99(1) | D31 | D22 | 1.4×a | 4.82(3) | ~0.17 |
| Add Charge #4 | 2.16(4) | D33 | D21 | 2.2×a | 2.14(4) | ~0.02 |
| Add Charge #7 | 2.22(5) | D32 | D11, D12, D13 | ~2×a | 2.19(5) | ~0.03 |
| Add Charge #11 | 8.06(1) | D33 | D23 | a | 6.74(3) | ~1.30 |

Table S4. Quantitative analysis of the charge addition energies from the avoided crossings in Figure S6 (c). In analogy to single-electron resonant tunneling in a double-dot system, the addition energy separation at an avoided crossing in a 3x3 array with two in-plane gates is characterized by $(U_m + 2t')$, where $U_m$ is the mutual charging energy (long-distance e-e repulsion) between the two charge addition distributions (see the red-blue charge distribution plots in Fig. S6(d)) that are on resonance at the avoided crossing, and $t'$ is the resonant tunnel coupling between them. The listed addition energies (2nd column) at the avoided crossings are extracted from the numerically simulated addition energy spectrum with hopping t=0.5meV, as shown in Figure S6(c). The $U_m$ at an avoided crossing is illustrated in the subplot in Figure S6(c). The $U_m$ values (6th column) at the avoided crossings are extracted from numerically simulated addigtion energy spectrym with hopping t=0meV. The uncertainty in the least significant digit is limited by the x-axis (gate voltage) resolution in the numerical simulation. The $2t'$ values in the last column are estimated as the difference between the addition energy and $U_m$ at the avoided crossings. The numbers support the observation that the mutual charging energy is inversely proportional to the spatial distance between the charge occupation configurations, while the resonant tunneling rate is exponentially dependent on this spatial distance.

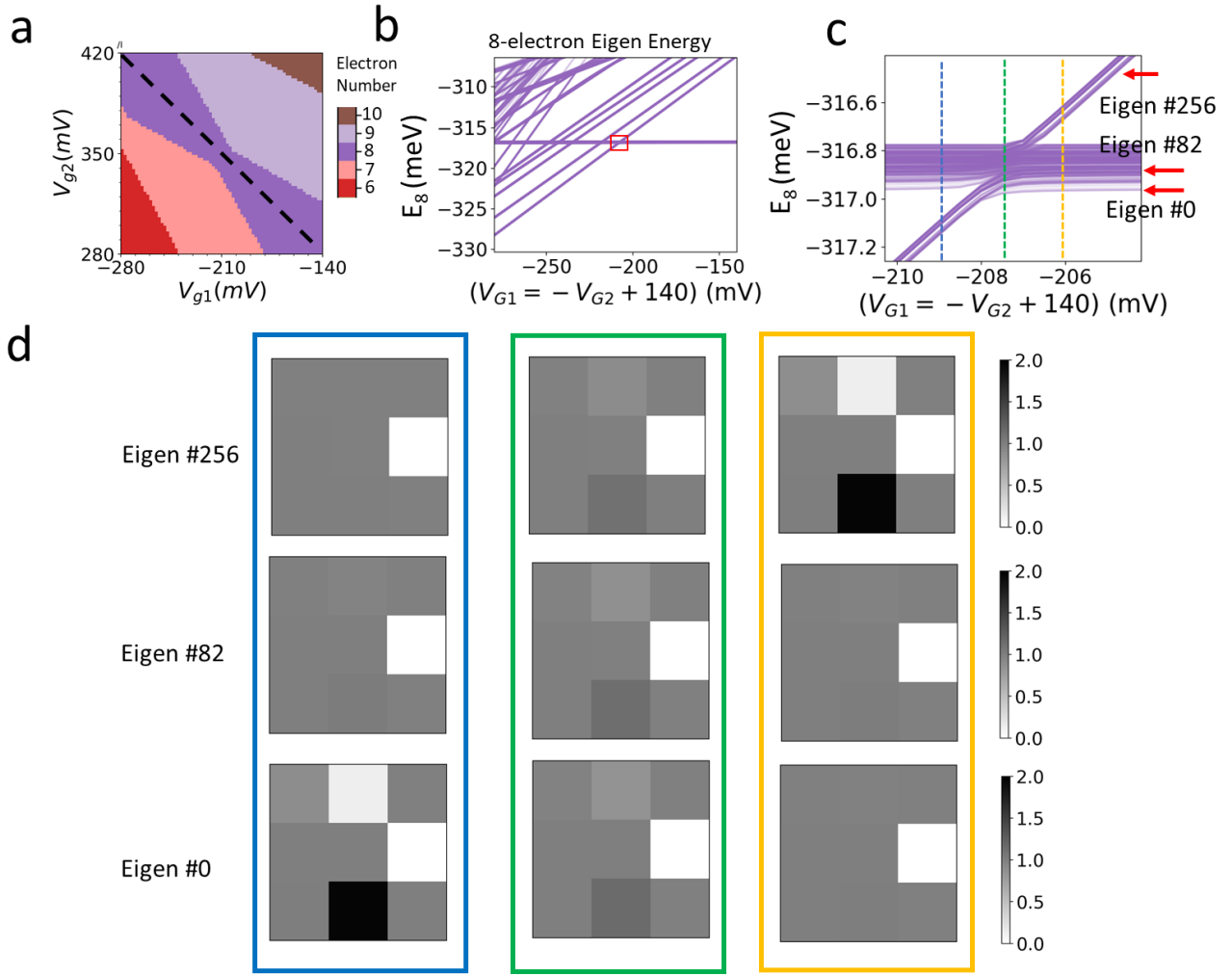

Figure S7. Eigen-energy spectrum and many-body charge distributions. (a) Avoided crossing region reproduced from Fig. 3a in the main text. (b) Simulated eigen-energy spectrum for occupation = 8 eigenstates along the dashed-line detuning axis in (a). (c) Close up eigen energy spectrum at the red circle in (b). The eigen energy lines are plotted with 80% transparency, and the color intensity reflects the level of degeneracy. The eigen energy levels bundle into minibands of highly degenerate or near-degenerate eigenstates. (d) The charge distribution of selected eigenstates at different locations along the detuning axis as highlighted in (c). Charge distribution configurations are nearly identical for eigenstates within the same miniband. Moving along the detuning axis, the ground state alters its charge distribution. The energy separation at the avoided crossing shown in (c) is determined by the tunnel coupling t′ between the two distributions whose eigenstates are hybridizing at the avoided crossing.

**S7. Characterizing Addition Energy Spectrum in the First and Second Arrays**

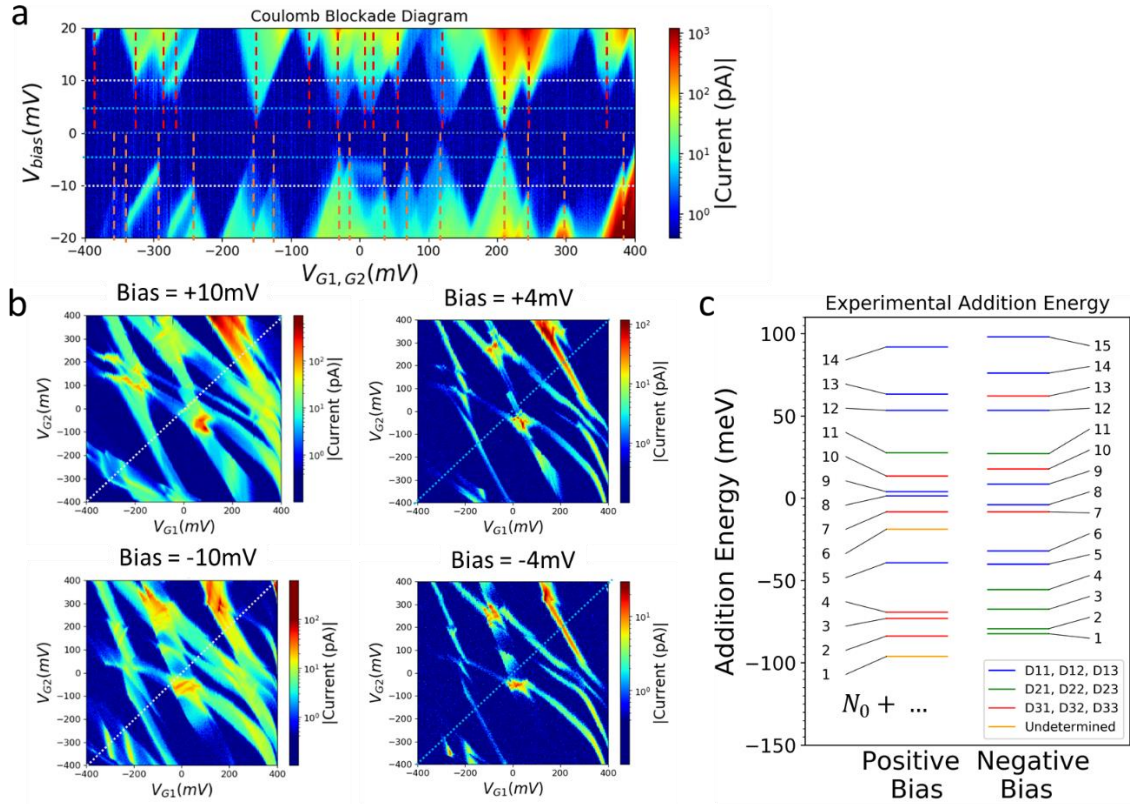

Figure S8. Extracting addition energy levels in the first array using the measured Coulomb blockade diagrams (a) Coulomb blockade conductance diagram taken along the 45-degree diagonal axis that is marked by dashed lines in (b). Horizontal dashed lines in (a) correspond to the bias voltage level of (b). The vertical dashed lines in (a) mark charge addition positions on the gate-voltage axis where conductance via the addition electrons becomes visible at finite biases. (b) Conductance gate-gate maps that are measured at representative biase voltages, +10mV, +4mV, -10mV, and -4mV, respectively. Maps are plotted in log-color scale to show conductance features of small amplitudes. (c) Charge addition energy spectra extracted by scaling the measured charge addition levels using properly chosen gate lever arms. The slopes of the charge addition lines in the gate-gate maps allow us to determine whether the conducting electron is added onto sites in the upper, middle, or lower row of the array, and to choose proper gate lever arms for converting charge addition positions on the gate voltage axis to energy levels in the addition energy spectrum. $N_0$ represents the initial number of excess electrons in the array.

Experimental determination of the charge addition spectrum in a quantum dot array is a prerequisite for more complex measurements and control of many-body states in the array. We use the positions of Coulomb oscillation peaks in the measured Coulomb blockade diagram, as marked by vertical dotted lines in Figure S8 (a), to extract the addition-energy spectrum along the 45-degree diagonal axis in the gate-gate voltage parameter space. To choose the proper gate lever arms to convert measured charge addition levels from the gate voltage axis to energy space, we determine the average gate lever arm $(\alpha_r = (\sum_{i \in r} \alpha_{G1,i} + \alpha_{G2,i})/3$ for adding an electron to sites in row $r$) according to the gate-gate slope of its charge addition boundary. We plot the charge addition energy spectra in Figure S8 (c), for both positive bias and negative bias conditions. The color of the energy levels corresponds to the gate-gate slope of the conduction lines, which also corresponds to the row of sites where the conducting electron is added. The differences in the extracted addition energy levels between a positive bias condition and a negative bias condition result from the differences in specific atomic configurations along the source/drain direction. We note that the absolute numbers of excess electrons in this device are unknown; however, the assigned charge numbers in Figure S8 (c) do not affect the underlying physical phenomena in this work.

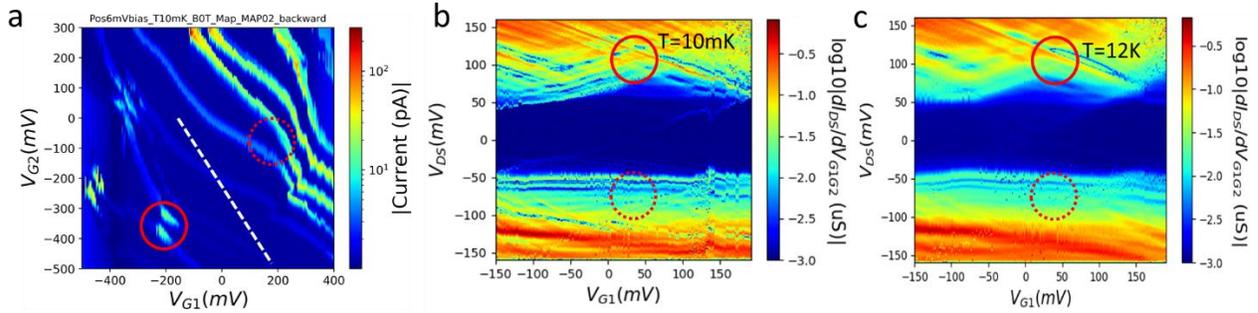

Figure S9. Characterizing the addition energy spectrum in the second array. (a) Gate-gate map of conductance through the array measured at $V_{bias} = 6mV$ and at the base temperature of the dilution refrigerator ($T = 10mK$). (b) (c) Finite bias differential conductance spectroscopy taken from the cut along the dashed line in (a), and measured at T=10mK (b) and T=12K (c), respectively. When doing a transport measurement, a large positive bias at the drain lead (with source lead grounded) pulls down the chemical potential in the drain lead, and individual electrons propagate through the lower Hubbard band within the bias window. Therefore, the differential conductance spectrum at positive (negative) biase in (b) represents the addition energy spectrum of the lower (upper) Hubbard band. The addition-energy levels in the measured lower Hubbard band section are detuned by the differential gate-gate voltages and exhibit crossings of addition energy levels (solid circle) in (b) and (c), which corresponds to the avoided crossing region (solid circle) in (a). In contrast, the addition-energy levels in the measured upper Hubbard band section are insensitive to the differential gate-gate voltages (dashed circles).